\documentclass[
aps,pra,
twocolumn
]{revtex4}

\usepackage{graphicx}
\usepackage{subfigure}
\usepackage{multirow}

\usepackage{amsmath}
\usepackage{amssymb}
\usepackage{amsthm}
\usepackage{braket}
\usepackage{color}
\usepackage{microtype}
\usepackage{tikz}
\usepackage{makecell}

\usepackage{hyperref}
\hypersetup{colorlinks,linkcolor={blue},citecolor={red},urlcolor={blue}}

\theoremstyle{plain}

\theoremstyle{plain}

\begin{document}

\title{A graph-based approach to  entanglement entropy of quantum error correcting codes}

\author{Wuxu Zhao}
\thanks{These two authors contribute equally}
\author{Menglong Fang}
\thanks{These two authors contribute equally}
\author{Daiqin Su}
\email{sudaiqin@hust.edu.cn}

\affiliation{National Gravitation Laboratory, MOE Key Laboratory of Fundamental Physical Quantities Measurement, Institute for Quantum Science and Engineering, and School of Physics, Huazhong University of Science and Technology, Wuhan 430074, People's Republic of China}

\date{\today}

\begin{abstract}
We develop a graph-based method to study the entanglement entropy of Calderbank-Shor-Steane quantum codes. This method offers a straightforward interpretation for the entanglement entropy of quantum error correcting codes through graph-theoretical concepts, shedding light on the origins of both the local and long-range entanglement.
Furthermore, it inspires an efficient computational scheme for evaluating the entanglement entropy. We illustrate the method by calculating the von Neumann entropy of subsystems in toric codes and two types of quantum low-density-parity check codes, and by comparing the scaling behavior of the entanglement entropy with respect to the subsystem size. Our method provides a new perspective for understanding the entanglement structure in quantum many-body systems.

\end{abstract}

\maketitle

{\it Introduction.} 
Entanglement plays a crucial role in understanding the physics of quantum many-body systems, including topological matters~\cite{Kitaev2006topological, Levin2006detecting, Eisert2010area}, the black hole information paradox~\cite{Hawking1976breakdown, Page1993information}, information scrambling in a thermalization process~\cite{Hayden2007black}, the connection between geometry and entanglement in quantum gravity~\cite{almheiri2015bulk, pastawski2015holographic}, and symmetry breaking/restoration~\cite{ares2023entanglement, Ares2024entanglement, liu2024symmetry}. In particular, the scaling law of the entanglement entropy with respect to subsystem size reveals key properties of quantum systems, e.g., discriminating critical and non-critical phases~\cite{Vidal2003entanglement, latorre2003ground}. Entanglement is introduced in quantum error correcting codes to encode logical information in a non-local manner and protect it from local errors~\cite{shor1995scheme}. This is manifested in their notable feature of local indistinguishability~\cite{knill1997theory, bennett1996mixed}. 
There exists a trade-off between the amount of entanglement and the code distance~\cite{bravyi2024much, li2024much}: the lower bound of the geometric entanglement is proportional to the code distance, which implies that more entanglement is required  to correct more errors. 

Given the unique role of entanglement in quantum error correcting codes, it is important to develop methods to calculate the amount of entanglement within quantum codes.
A group-theoretical method to evaluate the entanglement entropy for stabilizer codes was proposed by Refs.~\cite{fattal2004entanglement, hamma2005ground, Hamma2005bipartie}. It was used to study the topological entanglement entropy in toric codes~\cite{flammia2009topological} and other topological codes~\cite{kargarian2008entanglement}, and the entanglement dynamics in many-body systems~\cite{nahum2017quantum, li20219measurement}. Recently, a graph-based method was developed to study higher-rank topological phases~\cite{ebisu2023entanglement}. 

In this paper, we develop a different graph-based method to calculate the explicit form of the reduced density matrix and the entanglement entropy of Calderbank-Shor-Steane (CSS) quantum codes. It provides a clear picture on the origin of the local entanglement, which indicates the area law, and the long-range topological entanglement in the toric codes. Furthermore, it inspires an efficient algorithm to evaluate the entanglement entropy for all CSS codes, e.g., quantum low-density-parity-check (qLDPC) codes~\cite{tillich2013quantum, Kovalev2013quantum, Breukman2016,  panteleev2021degenerate, breuckmann2021balanced, panteleev2022asymptotically, leverrier2022quantum, higgott2023constructions, wang2023abelian, Lin2024quantum, Breuckman2021}. We demonstrate the graph-based method by calculating the entanglement entropy for the bivariate-bicycle codes~\cite{bravyi2024high} and quasi-cyclic codes~\cite{hagiwara2007quantum}, and studying their scaling law with respect to the size of the subsystem. 

{\it Two-dimensional toric code}. 
We briefly introduce some basic concepts in graph theory~\cite{diestel2024graph} that are essential for the subsequent discussions.
A graph $G(V,E)$ consists of a set of vertices $V$ and a set of edges $E$. An edge connecting two vertices $v_i,v_j \in V$ is represented by the pair $(v_i,v_j)$. A path is a sequence of distinct vertices connected by edges, and a cycle is a path that starts and ends at the same vertex. All cycles in a graph $G$ form a cycle space, the basis of which is the set of independent cycles. The cyclomatic number of $G$ is the count of these independent cycles~\cite{SI}. A connected graph is one in which there exists at least one path between any pair of vertices. A tree graph is a connected graph that contains no cycles. A spanning tree of a graph $G$ is a connected subgraph that includes all vertices without cycles.

\begin{figure}[htbp]
    \centering
    \includegraphics[width=0.95\linewidth]{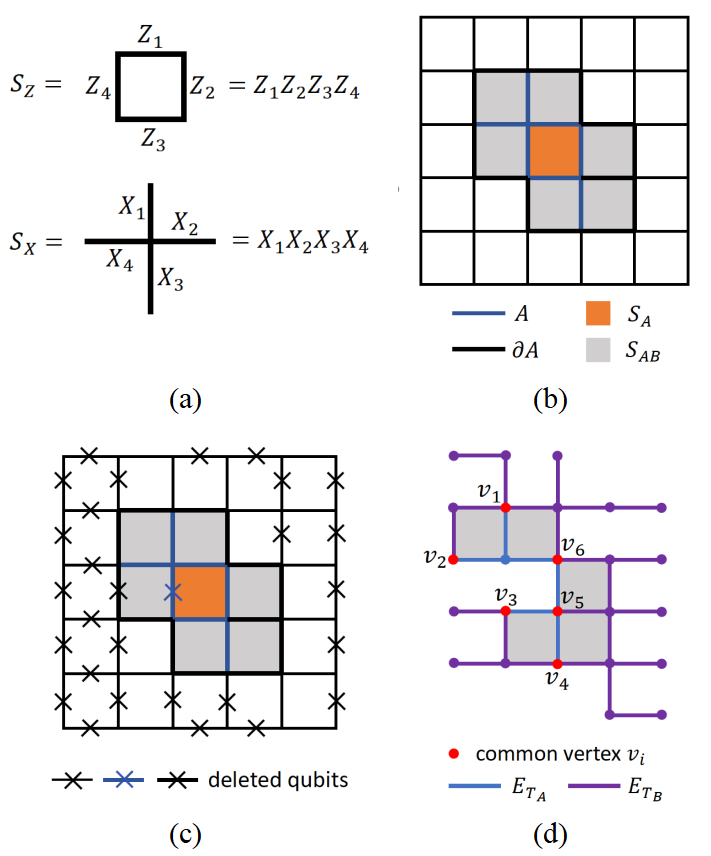}
    \caption{Illustration of subsystems and their spanning trees in a toric code embedded in a square lattice with periodic boundary conditions, initialized in the logical $Z$ eigenstate. (a) A $Z$-type stabilizer generator $S_Z$ acts on four qubits surrounding a face, while an $X$-type stabilizer generator $S_X$ acts on four qubits incident to a vertex. (b) Subsystem $A$ consists of qubits on blue edges and contains one cycle, highlighted by the orange face. Subsystem $B$ includes all remaining qubits on black edges. The exterior boundary of subsystem $A$ is indicated by the deep black edges. Stabilizers shared by subsystems $A$ and $B$ are marked by gray faces. (c) The spanning tree of subsystem $A$ is formed by removing one qubit (blue cross) from the orange face. The spanning tree of subsystem $B$ is constructed by removing qubits indicated by black crosses. (d) Blue edges $E_{T_A}$ indicate the spanning tree of subsystem $A$ and purple edges $E_{T_B}$ mark the spanning tree of subsystem $B$. The gray faces represent the joint cycles and the red vertices labeled by $v_i$ denote the common vertices of the two spanning trees. The entanglement entropy equals $5$, which is the number of independent joint cycles, as determined by $|\mathcal{C}_{T_A\cup T_B}|$ or computed via Eq.~\eqref{eq:graphic-form}.}
    \label{fig:region common}
\end{figure}

Two-dimensional toric codes can be represented by simple two-dimensional graphs, where qubits are placed on edges~\cite{kitaev2003fault}. The faces and vertices of the graph represent stabilizers of type $Z$ and $X$, respectively (see Fig.~\ref{fig:region common} (a)). There are two distinct sets of cycles in the graph, the contractible cycles and non-contractible cycles. The former corresponds to $Z$-type stabilizers, while the latter represents logical Pauli $Z$ operators. Since all stabilizers have an eigenvalue of one in the code subspace, the sum of qubit values along any contractible cycle is zero (mod 2), imposing a constraint on these qubit values. The sum of qubit values along a non-contractible cycle can be either 0 or 1, depending on the specific logical state. If one of the logical Pauli $Z$ operators takes a specific eigenvalue, the qubit values along the corresponding non-contractible cycle are also subject to a constraint.

We are concerned with the reduced density matrix and the von Neumann entropy of a subsystem $A$, a subset of qubits, of a toric code. The remaining qubits of the toric code are in the complementary subsystem $B$, as shown in Fig.~\ref{fig:region common} (b). There are cycles that are entirely enclosed by subsystem $A$ or subsystem $B$, as well as cycles shared by both subsystems, which are referred to as boundary stabilizers. 
The exterior boundary of subsystem $A$, denoted as $\partial A$, is defined as the set of edges (qubits) that belong to the boundary stabilizers but are not part of subsystem $A$.

Denote by $\mathcal{C}(A)$ and $\mathcal{C}(B)$ the sets of independent cycles in subsystems $A$ and $B$. By removing one edge from every independent cycle, we obtain spanning trees $T_A$ and $T_B$ within subsystems $A$ and $B$, respectively (see Fig.~\ref{fig:region common} (c)). The union of these two spanning trees $T_A$ and $T_B$ forms a graph $T_A \cup T_B$, which contains a set of independent joint cycles $\mathcal{C}(T_A \cup T_B)$ (see Fig.~\ref{fig:region common} (d)). The sum of qubit values along any independent joint cycle $c_k \in \mathcal{C}(T_A \cup T_B)$ is zero (mode 2), which implies 
\begin{eqnarray}\label{eq:joint-constraints}
    \sum_{e \in E_{c_k}^A} z_e=\sum_{e \in E_{c_k}^B}z_e=C_{\alpha}^k ~~ (\text{mod} \,\, 2).
\end{eqnarray}
Here, $z_e = (1-\lambda)/2 \in \{ 0, 1 \}$ is the qubit value on the edge $e$, with $\lambda = \pm 1$ the eigenvalue of Pauli $Z$ operator; and $E_{c_k}^A$ and $ E_{c_k}^B$ represent the edge (qubit) sets of the partitions in the two subsystems; and $C_{\alpha}^k \in \left\{0,1\right\}$, with $\alpha$ labeling a bit string that represents the values of the sums in Eq.~\eqref{eq:joint-constraints}.

For a given set of $\left\{C_{\alpha}^k\right\}$ (with $\alpha$ fixed), the $|\mathcal{C}(T_A \cup T_B)|$ constraints given by Eq.~\eqref{eq:joint-constraints} impose $|\mathcal{C}(T_A \cup T_B)|$ constraints on the qubit values in the spanning tree of subsystem $A$. Since there are $|E_{T_A}|$ qubits in the spanning tree $T_A$, the remaining $|E_{T_A}|-|\mathcal{C}(T_A \cup T_B)|$ qubits are independent. The qubit configurations of subsystem $A$ span a subspace $\Lambda^A_{\alpha}$ with dimension $d^A_{\alpha}=2^{|E_{T_A}|-|\mathcal{C}(T_A \cup T_B)|}$. The state of the subsystem $A$ is then an equal superposition of all allowable qubit configurations,
\begin{eqnarray}
    |\psi_{\alpha}\rangle=\frac{1}{\sqrt{d^A_{\alpha}}}\sum_{z~\in \Lambda^A_{\alpha}}|z_1, z_2,\cdots,z_{|E_A|}\rangle.
\end{eqnarray}
Similarly, for a given set of $\left\{C_{\alpha}^k\right\}$ (with $\alpha$ fixed), the $|\mathcal{C}(T_A \cup T_B)|$ constraints given by Eq.~\eqref{eq:joint-constraints} impose $|\mathcal{C}(T_A \cup T_B)|$ constraints on the qubits values on the spanning tree of the subsystem $B$. Since there are $|E_{T_B}|$ qubits in the spanning tree $T_B$, the remaining $|E_{T_B}|-|\mathcal{C}(T_A \cup T_B)|$ qubits are independent. The qubit configurations of the spanning tree $T_B$ also span a subspace $\Lambda^B_{\alpha}$ with the dimension $d^B_{\alpha}=2^{|E_{T_B}|-|\mathcal{C}(T_A \cup T_B)|}$. The pair of subspaces $(\Lambda^A_{\alpha},\Lambda^B_{\alpha})$ is uniquely determined by the set $\{ C^k_{\alpha} \}$ with $\alpha$ fixed.

For a different set of $\left\{C_{\alpha'}^k\right\}$, it follows from Eq.~\eqref{eq:joint-constraints} that the subspaces $\Lambda^B_{\alpha}$ and $\Lambda^B_{\alpha'}$ do not overlap and are therefore orthogonal. Since there are $|E_{T_B}|-|\mathcal{C}(T_A \cup T_B)|$ independent qubits in the spanning tree $T_B$, and the subspaces $\Lambda^B_{\alpha}$ and $\Lambda^B_{\alpha'}$ are of the same size, the number of distinct pairs of subspaces $(\Lambda^A_{\alpha},\Lambda^B_{\alpha})$ is $n_{\Lambda}=2^{|\mathcal{C}(T_A \cup T_B)|}$. After tracing out the degrees of freedom of subsystem $B$, we obtain the density matrix of subsystem $A$,
\begin{eqnarray}\label{eq:DM}
    \rho_A=\frac{1}{n_{\Lambda}}\sum_{\alpha=1}^{n_{\Lambda}}|\psi_{\alpha}\rangle\langle\psi_{\alpha}|.
\end{eqnarray}
The density matrix $\rho_A$ is in a diagonal form, so it is straightforward to calculate the von Neumann entropy, which is given by 
\begin{eqnarray}
    S_A=-\log_2(n_{\Lambda})=|\mathcal{C}(T_A \cup T_B)|.
\end{eqnarray}
This result highlights a direct graph-theoretic interpretation of entanglement: it quantifies the independent joint cycles formed by combining the spanning tree structures of separated subsystems.

The entanglement entropy can be expressed in terms of the number of connected components within subsystems $A$ and $B$, as well as the number of common vertices shared between them~\cite{SI}. 
\begin{eqnarray}\label{eq:graphic-form}
    S_A=|\mathcal{C}(T_A\cup T_B)|=|V_{A\cap B}| - K_1-K_2+1,
\end{eqnarray}
where $A \cap B$ is the intersection between $A$ and $B$, with $V_{A \cap B}$ the set of their common vertices, and $K_1$ and $K_2$ are the numbers of connected components within $A$ and $B$, respectively. This implies that the entanglement entropy depends on the global connectivity between subsystems, as well as the fragmentation within each subsystem.

When subsystem $A$ is a single connected component and its boundary consists of only one cycle, its von Neumann entropy is $S_A = m-1$, with $m$ the number of boundary stabilizers. The term $-1$ is associated with the cycle surrounding subsystem $A$ and is independent of the size of subsystem, indicating a pattern of long-range entanglement. The entanglement entropy can also be evaluated for the case where the boundary encompasses multiple cycles. We explicitly calculate the von Neumann entropy for several types of subsystem within a toric code~\cite{SI}.

{\it General formalism for CSS codes}. 
A general CSS code~\cite{calderbank1996good, steane1996multiple} cannot always be represented by a simple graph like toric and surface codes. However, we can still leverage related concepts in graph theory (such as cycles and spanning trees) and generalize the above graph-based method to calculate the density matrix and von Neumann entropy of a subsystem in a CSS code, using its parity check matrix $H_Z$.
This is because the multiplication of stabilizer generators can be represented by the addition (mod 2) of row vectors in the parity check matrix $H_Z$, and the latter is equivalent to the symmetric difference in the cycle space~\cite{SI}. 
The $Z$-type stabilizer generators of CSS codes can be partitioned into three sets: (1) $s_A$ generators acting only on qubits within subsystem $A$, denoted as $\big\{S_A^{(i)} \big\}_{i=1}^{s_A}$;
(2) $s_B$ generators acting only on qubits within subsystem $B$, denoted as $\big\{S_B^{(i)} \big\}_{i=1}^{s_B}$;
and (3) $s_{AB}$ generators acting on qubits in both subsystems $A$ and $B$, denoted as $\big\{S_{AB}^{(i)}\big\}_{i=1}^{s_{AB}}$.

For a specific stabilizer generator $S_A^{(i)}$ that acts only on qubits in subsystem $A$, one can delete one of its qubits to remove the cycle. 
Suppose that we want to delete one of the qubits that the stabilizer $S_A^{(i)}$ acts on. If another stabilizer generator, say $S$, also acts on this qubit, meaning that its element in the corresponding column is also one, we then replace the stabilizer generator $S$ with a new stabilizer generator $S' = S_A^{(i)} S$, for which the element in the corresponding column becomes zero. We continue this process to replace all other stabilizer generators acting on this qubit with new stabilizer generators. The parity check matrix is then transformed into a form in which this qubit is acted on only by the stabilizer generator $S_A^{(i)}$. We delete the qubit and the stabilizer $S_A^{(i)}$ from the parity check matrix. The remaining matrix is still a valid parity check matrix, with its rows representing a different set of stabilizer generators. The procedure described above is equivalent to Gaussian elimination. 
We continue the above procedure to delete all stabilizer generators acting only on qubits in subsystem $A$ and those acting only on qubits in subsystem $B$. The parity matrix is transformed into a form like $M_{AB} = \begin{pmatrix}
 M_A & M_B \\
\end{pmatrix}$, which describes the stabilizer generators shared by subsystems $A$ and $B$.

The boundary of subsystem $A$ may contain one or more cycles, which can be generated by multiplying several stabilizer generators shared between $A$ and $B$. This implies that new stabilizer generators can be defined from $M_{AB}$, which act only on qubits on the boundary of subsystem $A$. Similarly, if additional cycles exist within subsystem $A$, new stabilizer generators acting exclusively on qubits in $A$ can be defined from $M_{AB}$. 
Once all such generators have been identified,
we apply the above procedure to remove those cycles~\cite{SI}.

By reintroducing the deleted qubits and stabilizer generators, the original parity check matrix $H_Z$ can be transformed into 
$\tilde{H}_Z$ through Gaussian elimination, along with appropriate row and column permutations,
 \begin{eqnarray}
 \tilde{H}_Z = 
    \begin{pmatrix}
         \mathbb{I} & \tilde{W}_A & \boldsymbol{0} & \boldsymbol{0} \\
         \boldsymbol{0} & W_A & W_B & \boldsymbol{0} \\
         \boldsymbol{0} & \boldsymbol{0} & \tilde{W}_B & \mathbb{I}
     \end{pmatrix}.
 \end{eqnarray}
Define the qubit configuration of subsystem $A$ as $|z_A \rangle = |z_A^d, z_A^t \rangle $ and that of subsystem $B$ as $|z_B\rangle =|z_B^t, z_B^d \rangle$. Here, $|z_A^d \rangle$ and $\ket{z_A^t}$ represent the qubit configurations of the deleted and remaining qubits in subsystem $A$, respectively; while $|z_B^d \rangle$ and $\ket{z_B^t}$ denote the qubit configurations of the deleted and remaining qubits in subsystem $B$, respectively. 
The qubit configuration $\ket{z} = \ket{z_A, z_B}$ satisfies $\tilde{H}_Z \, z = 0$, which implies
 \begin{eqnarray}\label{eq:solution-constraint}
     &z_A^d + \tilde{W}_A z_A^t=0, 
     ~~~~
     z_B^d + \tilde{W}_B z_B^t=0, & \nonumber \\
     & W_A z_A^t = W_B z_B^t = C_{\alpha},
 \end{eqnarray}
 where $C_{\alpha}$ is a binary vector whose number of components equals to the number of rows of $W_A$ and $W_B$. 
 
 Suppose that the number of remaining qubits in subsystem $A$ is $t_A$ and that in subsystem $B$ is $t_B$, and the rank of the submatrices $W_A$ and $W_B$ is $r$. 
 For a given $C_{\alpha}$, there are $2^{t_A-r}$ qubit configurations $|z_A \rangle$ that satisfy the constraints in Eq.~\eqref{eq:solution-constraint}. Consequently, these allowable configurations form a subspace $\Lambda_{\alpha}^A$ with dimension $d_{\alpha}^A=2^{t_A-r}$. 
 Similarly, there are $2^{t_B-r}$ qubit configurations $\ket{z_B}$ for the same $C_{\alpha}$, which form a subspace ${\Lambda}_{\alpha}^B$ with dimension ${d}_{\alpha}^B=2^{t_B-r}$. 
 Different binary vectors $C_{\alpha}$ define mutually orthogonal subspaces for subsystems $A$ and $B$, which implies that the number of pairs of orthogonal subspaces $(\Lambda_{\alpha}^A, \Lambda_{\alpha}^B)$ is $n_{\Lambda}=2^r$. After tracing out the degrees of freedom of subsystem $B$, the density matrix of subsystem $A$ is given in the same form as Eq.~\eqref{eq:DM}, with $n_{\Lambda}=2^r$.
One can now straightforwardly calculate the von Neumann entropy, 
\begin{eqnarray}
  S_A=-\log_2(n_{\Lambda})=r=\text{rank}(W_A)=\text{rank}(W_B).
\end{eqnarray}
Note that the submatrix $\begin{pmatrix}
    W_A&W_B
\end{pmatrix}$ generates the joint cycle space formed by the spanning trees of $A$ and $B$, therefore the entanglement entropy is equal to the cyclomatic number of the joint cycle space.

For a given parity check matrix $H_Z=\begin{pmatrix}
    H_A & H_B \\
\end{pmatrix}$, for which $\text{rank}(H_A)=r_A$, $\text{rank}(H_B)=r_B$ and $\text{rank}(H_Z)=r_H$, it can be proved  that~\cite{SI}
\begin{eqnarray}\label{eq:General-formalism-for-CSS-codes}
    S_A = r_A + r_B - r_H.
\end{eqnarray}
This provides an efficient scheme for evaluating the entanglement entropy, since the rank of a matrix can be calculated efficiently. The entanglement entropy for other logical states can be evaluated similarly by appending rows representing logical Pauli $Z$ operators to the parity check matrix $H_Z$~\cite{SI}.

{\it Entanglement entropy for qLDPC codes}. 
We apply the graph-based method to calculate the entanglement entropy for two families of qLDPC codes: bivariate bicycle (BB) codes~\cite{bravyi2024high} and quasi-cyclic (QC) codes~\cite{hagiwara2007quantum}, both of which are quantum codes of CSS type. In BB codes, each stabilizer generator acts non-trivially on six qubits and each qubit participates in six stabilizer generators; while QC codes exhibit variable stabilizer weights. Neither BB codes nor QC codes are geometrically local. We show that the entanglement properties of the BB codes and QC codes differ significantly from those of the two-dimensional toric codes~\cite{SI}.

\begin{figure}
{\includegraphics[width=0.85 \columnwidth]{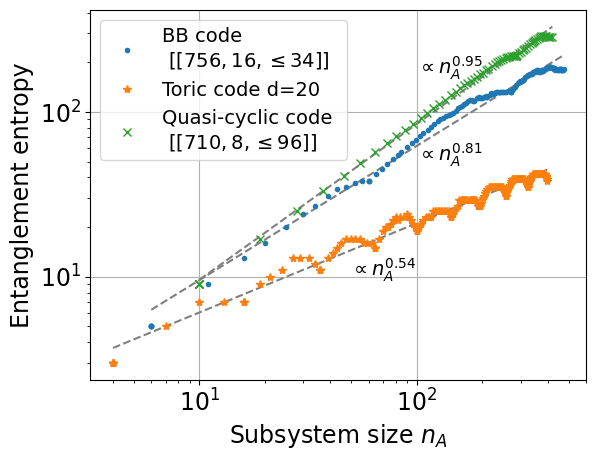}}
\caption{ Scaling of entanglement entropy for toric codes, bivariate bicycle codes and quasi-cyclic codes. The size of the subsystem is chosen to be less than half of the entire system. }
\label{fig:entropy-scaling}
\end{figure}

The entanglement entropy for a toric code with code distance $d=20$, a BB code $[[756, 16, \le 34]]$ and a QC code $[[710, 8, \le 96]]$
is shown in Fig.~\ref{fig:entropy-scaling}. In this calculation, the subsystem $A$ is selected in a specific way such that it remains a single connected component as its size increases~\cite{SI}. The entanglement entropy for the toric code is approximately proportional to $\sqrt{n_A}$, consistent with an area law that reflects the localized nature of its entanglement. In contrast, for the BB and QC codes, the entanglement entropy scales as $n_A^{\gamma}$ with $\gamma \approx 0.81$ and $0.95$, respectively, showing a faster growth in the entanglement entropy and revealing the presence of more delocalized and long-range entanglement.

\begin{figure*}[htbp]
  \centering
  \subfigure[]
  {\includegraphics[width=0.67 \columnwidth]{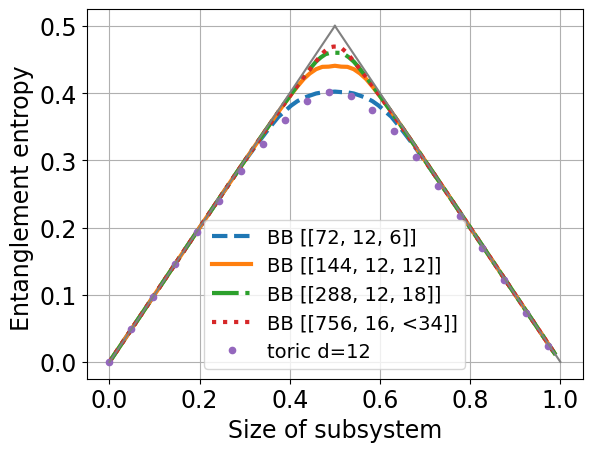}}
  \subfigure[] {\includegraphics[width=0.695 \columnwidth]{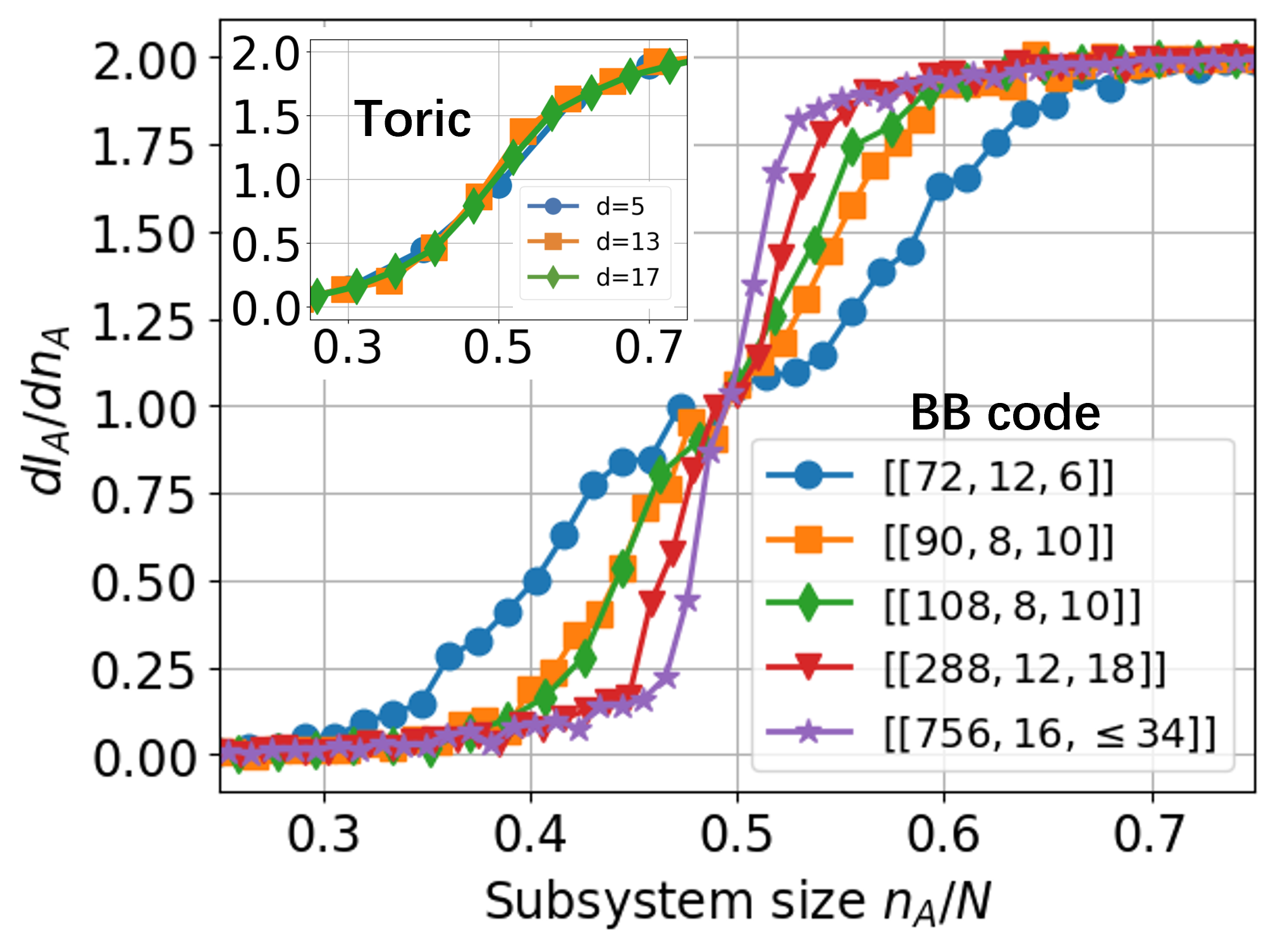}}
  \subfigure[]{\includegraphics[width=0.698 \columnwidth]{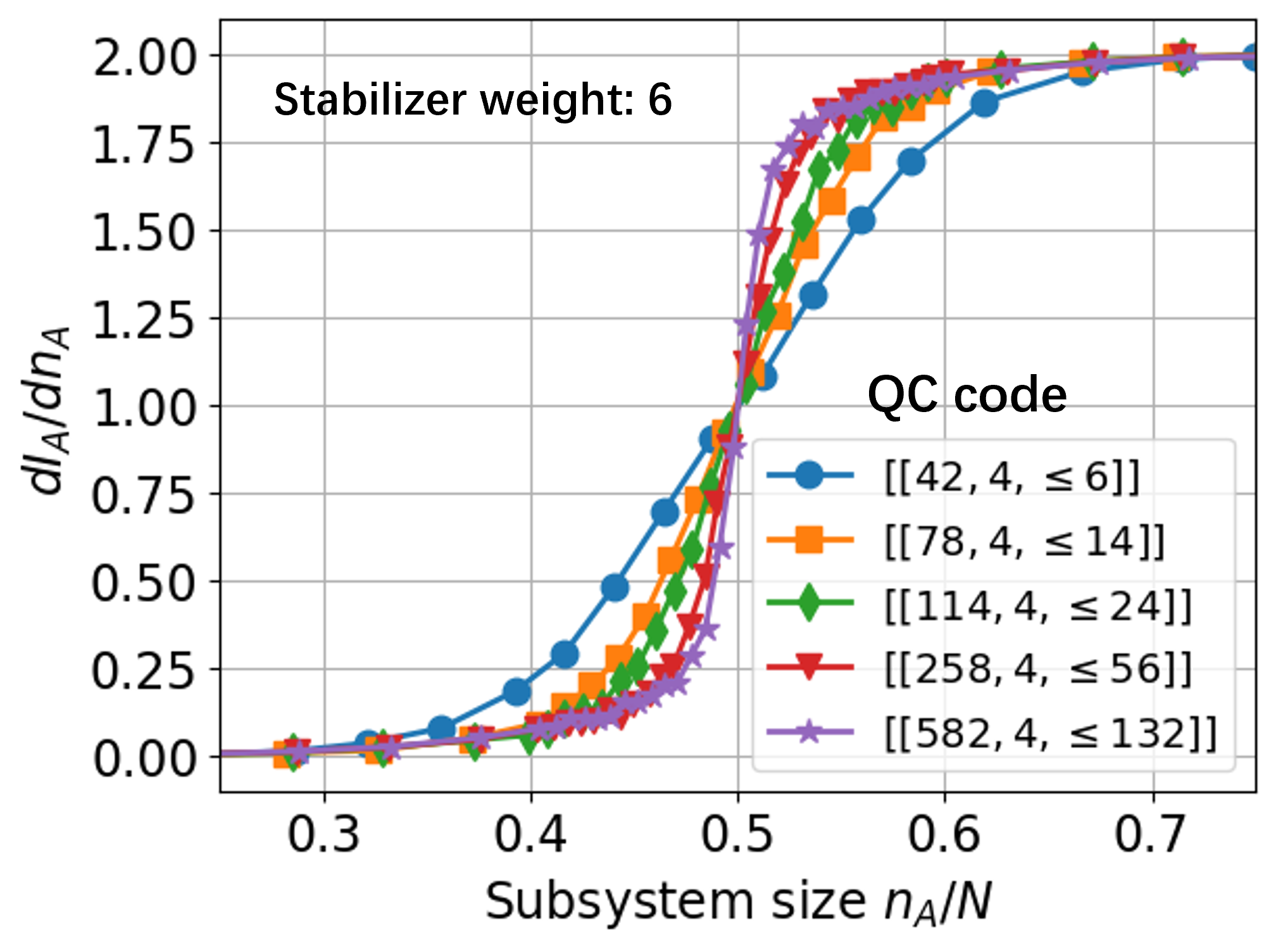}}
  \subfigure[]
  {\includegraphics[width=0.7 \columnwidth]{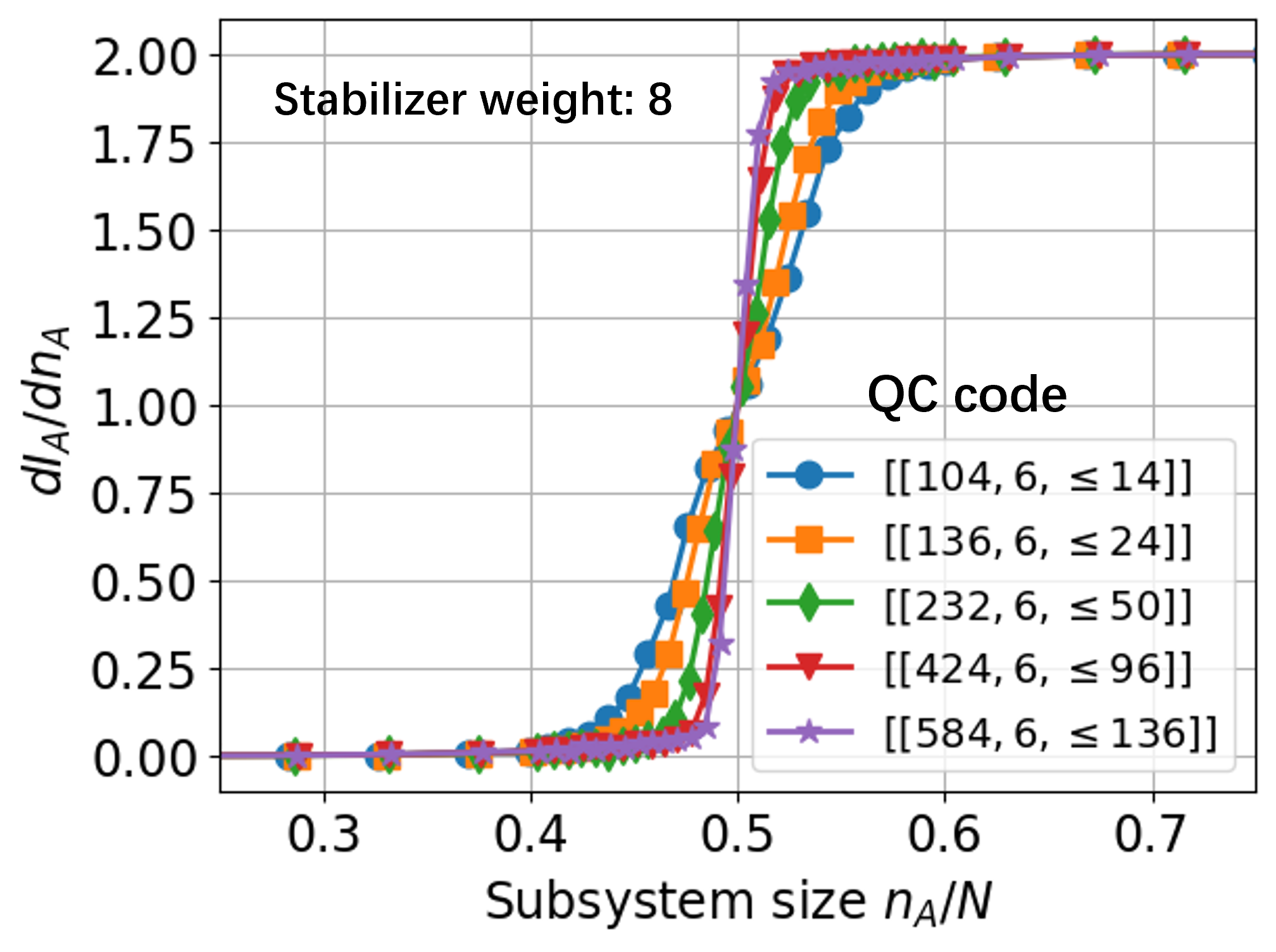}}
\caption{ Entanglement entropy for qLDPC codes and toric codes. (a) Average entanglement entropy for randomly selecting subsystems. (b) Derivative of the discrepancy of entropy $I_A$ for BB codes. The inset shows the same quantity for toric codes. (c) Derivative of $I_A$ for QC codes with stabilizer weight 6. (d) Derivative of $I_A$ for QC codes with stabilizer weight 8. }
\label{fig:entropy}
\end{figure*}

We then randomly select subsystems $A$ and calculate their average von Neumann entropy $\Bar{S}_A$, also called the average entanglement entropy. The results for toric codes and BB codes are shown in Fig.~\ref{fig:entropy} (a). It can be seen that the average entanglement entropy for sufficiently small random subsystems follows a volume law. This is due to the unique characteristics of the stabilizer codes. When a randomly selected subsystem is small compared to the entire system, the selected quibts are likely to be disconnected from each other, forming many isolated qubit ``islands". The state of the qubit island is maximally mixed, and its entanglement entropy is typically equal to its size, leading to the observed volume law. As the size of a randomly selected subsystem grows such that a qubit island encompasses one or more stabilizer generators (cycles in the language of graph theory), the average entanglement entropy starts to deviate from the volume law. We define the discrepancy between subsystem size $n_A$ and $\Bar{S}_A$ as a quantitative measure to characterize the deviation, $I_A = n_A - \Bar{S}_A (n_A)$.
The entropy discrepancy $I_A$ depends on the number of independent stabilizer generators that act exclusively within subsystem $A$.

We calculate the rate of change of the entropy discrepancy, ${\rm d} I_A/ {\rm d} n_A$, for both toric codes and qLDPC codes, revealing a significant difference between the two, as shown in Figs.~\ref{fig:entropy} (b)-(d). A sharp transition from a phase with a vanishing rate to one with a constant rate is observed for the qLDPC codes, with the transition becoming sharper as the size of the codes increases. In contrast, for the toric codes, the transition is much smoother and the steepness of the transition is independent of the size of the toric codes. The difference can be qualitatively attributed to variations in connectivity. Toric codes are defined on two-dimensional surfaces and exhibit strictly local interactions. As a result, the connectivity between subsystems, which is directly related to the entanglement entropy (see Eq.~\eqref{eq:graphic-form}), remains independent of the code size. In contrast, the high stabilizer weights and non-local interactions in BB and QC codes enhance connectivity between subsystems, leading to distinct entanglement properties. This is also evident, by comparing Figs.~\ref{fig:entropy} (c) and (d), from the observation that QC codes with higher stabilizer weights display a sharper transition.

The sharp transition in the rate of change of the entropy discrepancy in BB and QC codes reflects a distinct mechanism compared to toric codes. 
When subsystem $A$ is much smaller than half of the entire code and subsystem $B$ is a single connected component, transferring one qubit from $B$ to $A$ increases the entropy discrepancy only when a cycle or an $X$-type stabilizer is formed in $A$~\cite{SI}, the likelihood of which depends on both the stabilizer weight and the connectivity of the code. A larger stabilizer weight implies that subsystem $A$ must grow to a greater size before a cycle or an $X$-type stabilizer can be formed via adding a qubit, leading to a sharper transition. Due to the local nature of the toric code, its connectivity remains invariant as the code size increases, giving rise to ${d I_A}/{d n_A}$ that is independent of the code size. In contrast, for qLDPC codes, the connectivity evolves with the code size such that qubits participating in a stabilizer become more spatially separated (if embedded in a two-dimensional plane), rendering the formation of cycles or $X$-type stabilizers in $A$ increasingly difficult. This behavior serves as a clear empirical signature of the fundamental difference in entanglement properties between locally and nonlocally constrained quantum codes.

{\it Conclusions}. 
We developed a graph-based method to explore the density matrix and the entanglement entropy of CSS-type quantum error correcting codes. The entanglement entropy has a simple graph-theoretical interpretation: it is the cyclomatic number of the cycle space formed by joining the spanning trees of the subsystem and its complementary subsystem.
Its analytic expression related to the matrix rank inspires an efficient scheme to calculate the entanglement entropy for complicated quantum codes, e.g., qLDPC codes. We demonstrated the method by calculating and comparing the entanglement entropy for the toric codes and two families of qLDPC codes: the BB and QC codes, revealing significantly distinct entanglement properties between these two sets of quantum codes. These entanglement properties could be connected to the information protection capacity and decoding performance of CSS codes, which warrant further investigation in future work. The graph-based method does not depend on the specific geometry and topology of the quantum systems and can be generalized to explore systems with higher spins and irregular structures, as well as a broader class of stabilizer codes.

{\bf Acknowledgements:}D. S. is supported by the Fundamental Research Funds for the Central Universities, HUST (Grant No. 5003012068) and Wuhan Young Talent Research Funds (Grant No. 0106012013).

\bibliography{ref_EE_QEC}

\onecolumngrid
\appendix
\vspace{0.5cm}

\section{Edge space, cycle space and symmetric difference}\label{app:cycle space}

One can extract a vector space from a graph. The process involves extracting the edge set from a graph, which subsequently allows for the construction of a corresponding vector space based on that set. In graph theory, this space is referred to as the edge space~\cite{diestel2024graph}. Before constructing the vector space, it is essential to introduce the concept of symmetric difference between two sets. Given two sets $A$ and $B$, the symmetric difference between $A$ and $B$ is the set of elements that are in either $A$ or $B$, but not in their intersection. This is denoted as $A \Delta B$ and can be expressed mathematically as
\begin{eqnarray}
    A \Delta B = (A \setminus B) \cup (B \setminus A),
\end{eqnarray}
where the $A \setminus B$ is a set which contains the elements in $A$ but not in $B$.

The edge space comprises the subsets of $E$, where vector addition is defined by the symmetric difference. The basis of the edge space consists of all individual edges, which implies that the dimension of the edge space is equal to the size of the graph. We can map the edge space to a vector space $\Gamma$ over the field $\mathbb{F}_2$ whose dimension is the same as the size of the graph. For an edge subset within the edge space, we suppose that it is mapped to the vector $\tau_1$ in the space $\Gamma$. If a specific edge is absent, the corresponding component in $\tau_1$ is zero; while if the edge is present, the corresponding component is one. As a result, each edge subset can be mapped to a binary vector with dimension of $|E|$. The symmetric difference of two edge subsets corresponds to the addition (mod 2) between two binary vectors in $\Gamma$.
\begin{figure}
    \centering
    \includegraphics[width=0.4\linewidth]{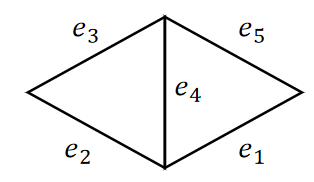}
    \caption{Illustration of the edge space and cycle space in a graph. Consider a given graph, where a basis of the edge space is comprised of the edges $e_1,e_2,e_3,e_4,e_5$. If we consider the edge subsets in $\text{pow}(E)$, $\omega_1=\left\{e_1,e_4,e_5\right\}$ and $\omega_2 = \left\{e_2,e_3,e_4\right\}$, the symmetric difference of the two edge subsets is given by $\omega_1 \Delta \omega_2 =\left\{e_1,e_2,e_3,e_5\right\}$. This operation demonstrates the combination of edges resulting from the symmetric difference, which is a key concept in understanding the edge space and its algebraic structure. The corresponding vectors for $\omega_1$ and $\omega_2$ in $\Gamma$ are $(1,0,0,1,1)$ and $(0,1,1,1,0)$, respectively. $\omega_1 \Delta \omega_2$ corresponds to the addition (mod 2) of $(1,0,0,1,1)$ and $(0,1,1,1,0)$. $(1,0,0,1,1)+(0,1,1,1,0)=(1,1,1,0,1)$ means that $\omega_1 \Delta \omega_2 = \left\{e_1,e_2,e_3,e_5\right\}$. The cycle space contains three cycles $\left\{e_1,e_4,e_5\right\},\left\{e_2,e_3,e_4\right\},\left\{e_1,e_2,e_3,e_5\right\}$. The dimension of the cycle space is 2.}
    \label{fig:enter-label}
\end{figure}

The cycle space of a graph $G$ consists of all cycles in the graph, denoted as $c(G)$. It is a subspace of the edge space. The vector addition of cycle space also amounts to the symmetric difference. The basis of the cycle space is the minimal set of cycles which can span all the cycles in the graph, with the elements of this set being defined as independent cycles. The dimension of the cycle space is defined as the number of elements in the basis, which is called cyclomatic number. 

Consider a parity check matrix $H_Z$, the multiplication of two stabilizer generators is equivalent to the addition (mod 2) of two binary vectors of its row space. All the stabilizers are generated by multiplying several stabilizer generators in the parity check matrix. Therefore, we can regard all independent generators as a basis of a cycle space. Each stabilizer corresponds to a cycle in the cycle space and the qubits associated with the stabilizers can be regarded as the edges of the cycle.

\section{Multiple boundary cycles}\label{app:multiple-boundary}

It is possible that the boundary $\partial A$ divides a plane into $\ell +1$ regions, with $\ell$ internal regions that are bounded by the edges of $\partial A$ and are denoted as $\{ R_\mu \}_{\mu = 1}^\ell$.
In this case, there are only $|E_{\partial A}|-\ell$ independent qubits and $2^{|E_{\partial A}|-\ell}$ distinct qubit configurations in the boundary $\partial A$. Denote the boundary of these regions as $\{ \partial R_\mu \}_{\mu = 1}^\ell$. Suppose the overlap between the set of vertices of the spanning tree $T_A$ and that of the region boundary $\partial R_\mu$ is $V_{T_A} \bigcap V_{\partial R_\mu}=\left\{v_1^{(\mu)},\cdots,v_{m_\mu}^{(\mu)}\right\}$, where $m_\mu$ is the total number of common vertices and the index $\mu$ indicates which region boundary these common vertices belong to. The common vertices are ordered counterclockwise along the shortest closed walk along the region boundary $\partial R_\mu$ and $v_1^{(\mu)}$ can be chosen arbitrarily. Divide the region boundary $\partial R_\mu$ as subsets based on the common vertices as follows,
\begin{eqnarray}\label{eq:boundaryR-division}
    \partial R_\mu = \partial R_\mu^{v_1,v_2}\bigcup \partial R_\mu^{v_2,v_3} \bigcup \cdots 
    \bigcup \partial R_\mu^{v_{m_\mu},v_1},
\end{eqnarray} 
where $\partial R_\mu^{v_j,v_{j+1}}$ is the path on the region boundary $\partial R_\mu$ between the vertex $v_j^{(\mu)}$ and the vertex $v_{j+1}^{(\mu)}$. This path does not contain other common vertices. Note that we have ignored the index $\mu$ associated with $v_j^{(\mu)}$ in Eq.~\eqref{eq:boundaryR-division} because information about the region boundary has already been indicated by the lower index of $\partial R_\mu$.

There is only one path between the vertex $v_j^{(\mu)}$ and the vertex $v_{j+1}^{(\mu)}$ in the spanning tree $T_A$, denoted as $P_{A[\mu]}^{v_j,v_{j+1}}$.
Since one of the edges in $\partial R_\mu$ has been deleted, there are $(m_\mu - 1)$ pairs of $(\partial R_\mu^{v_j,v_{j+1}}, P_{A[\mu]}^{v_j,v_{j+1}})$ that form joint cycles. Define the sum of qubit values in the edge set $E_{\partial R_ \mu}^{v_j,v_{j+1}}$ of the path $\partial R_\mu^{v_j,v_{j+1}}$ as $S_{\partial R_\mu}^{v_j,v_{j+1}}$, and that in the edge set $E_{P_{A[\mu]}}^{v_j,v_{j+1}}$ of the path $P_{A[\mu]}^{v_j,v_{j+1}}$ as $S_{A[\mu]}^{v_j,v_{j+1}}$, namely,
\begin{eqnarray}
    S_{\partial R_\mu}^{v_j,v_{j+1}} &=& \sum_{(v,\bar{v})\in E_{\partial R_\mu}^{v_j,v_{j+1}}} z_{(v,\bar{v})}, \nonumber\\
    S_{A[\mu]}^{v_j,v_{j+1}} &=& \sum_{(v',\bar{v}')\in E_{P_{A[\mu]}}^{v_j,v_{j+1}}} z_{(v',\bar{v}')},
\end{eqnarray}
where $z_{(v,\bar{v})}$ and $z_{(v',\bar{v}')}$ represent qubit values in the edge sets $E_{\partial R_\mu}^{v_j,v_{j+1}}$ and $E_{P_{A[\mu]}}^{v_j,v_{j+1}}$, respectively. These boundary cycles give rise to $m_\mu - 1$ constraints, which can be expressed as
\begin{eqnarray}
    S_{\partial R_\mu}^{v_j,v_{j+1}}+S_{A[\mu]}^{v_j,v_{j+1}}=0
\end{eqnarray}
or
\begin{eqnarray}\label{eq:constraint2}
     S_{\partial R_\mu}^{v_j,v_{j+1}}=S_{A[\mu]}^{v_j,v_{j+1}}=C_{\alpha}^{\mu,j}
\end{eqnarray}
for all possible value of $j$, and $C_{\alpha}^{\mu,j} \in \{0, 1 \}$. Taking into account $\ell$ boundary regions, there are $\sum_{\mu = 1}^\ell m_\mu - \ell$ constraints in total. 

When the qubit values on the boundary $\partial A$ are fixed, constraints given by Eq.~\eqref{eq:constraint2} are reduced to $\sum_{\mu=1}^\ell m_\mu - \ell$ constraints to the qubit values in the spanning tree $T_A$. 
The allowable qubit configurations of subsystem $A$ span a subspace $\Lambda_{\alpha}^A$ of dimension $d^A_{\alpha}=2^{|E_{T_A}|-\sum_\mu m_\mu + \ell}$. The state of subsystem $A$ is an equal superposition of all these qubit configurations,
\begin{eqnarray}
    |\psi_{\alpha}\rangle= \frac{1}{\sqrt{d^A_{\alpha}}}\sum_{z \in \Lambda_{\alpha}^A} | z \rangle = \frac{1}{\sqrt{d^A_{\alpha}}}\sum_{z \in \Lambda_{\alpha}^A} |z_1, z_2, \cdots , z_{n_A}\rangle.
\end{eqnarray}
For all qubit configurations in $\Lambda_{\alpha}^A$, $S_{A[\mu]}^{v_j,v_{j+1}}$ are fixed for all values of $\mu$ and $j$. 
In this case, the qubit configurations of the boundary $\partial A$ that satisfy Eq.~\eqref{eq:constraint2} span a subspace $\Lambda_\alpha^B$ with dimension $d^B_{\alpha}=2^{|E_{\partial A}|-\sum_\mu m_\mu}$. 
The mapping between the subspaces $\Lambda_{\alpha}^A$ and $\Lambda_{\alpha}^B$ is one-to-one, since they are determined by a fixed set of values $S_{A[\mu]}^{v_j,v_{j+1}}$.

For a different set of values $C_{\alpha'}^{\mu,j}$, it is evident from Eq.~\eqref{eq:constraint2} that the subspaces $\Lambda_{\alpha}^B$ and $\Lambda_{\alpha'}^B$ have no overlap and are therefore orthogonal. Since there are $|E_{\partial A}|-\ell$ independent qubits before joining the spanning trees $T_A$ and $T_B$, and the subspaces $\Lambda_{\alpha}^B$ and $\Lambda_{\alpha'}^B$ have the same size, the number of different pairs of subspaces $(\Lambda_{\alpha}^A,\Lambda_{\alpha}^B)$ is
\begin{eqnarray}
    n_{\Lambda} = \frac{2^{|E_{\partial A}|-\ell}}{d_\alpha^B} = \frac{2^{|E_{\partial A}|-\ell}}{2^{|E_{\partial A}|-\sum_\mu m_\mu}} = 2^{\sum_\mu m_\mu-\ell}.
\end{eqnarray}
After tracing out the degrees of freedom of subsystem $B$, we obtain the density matrix of subsystem $A$,
 \begin{eqnarray}
     \rho_A=\frac{1}{n_{\Lambda}} \sum_{\alpha=1}^{n_{\Lambda}} |\psi_{\alpha}\rangle \langle\psi_{\alpha}|.
 \end{eqnarray}
 The density matrix $\rho_A$ is in a diagonal form, so it is straightforward to calculate the von Neumann entropy, which is equal to \begin{eqnarray}
     S_A = -\log_2(n_{\Lambda})=\sum_{\mu=1}^\ell m_\mu -\ell.
 \end{eqnarray}
 
The edges in every pair of subgraphs $(\partial R_\mu^{v_j,v_{j+1}}, P_{A[\mu]}^{v_j,v_{j+1}})$ constitute one joint cycle. One needs to delete one edge in each region boundary cycle. 
If the deleted edge lies at the intersection of two joint cycles, the two faces formed by the two joint cycles become connected. In another case, if the deleted edge is incident with both the face formed by a joint cycle and the external region, the two regions become connected. In both cases, deleting an edge reduces one face formed by the joint cycles. Before the process of deleting an edge in each region boundary, there are a total of $\sum_\mu m_\mu$ faces formed by the joint cycles.  After deleting $\ell$ edges, the number of remaining faces is $\sum_\mu m_\mu - \ell$, which corresponds to the entanglement entropy.

\section{Some examples in toric codes}\label{app:example}

\begin{figure}[htbp]
  \centering
  \subfigure[]
  {\includegraphics[width=0.20 \columnwidth]{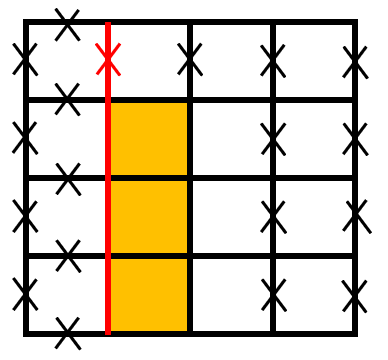}}
  \subfigure[]
  {\includegraphics[width=0.20 \columnwidth]{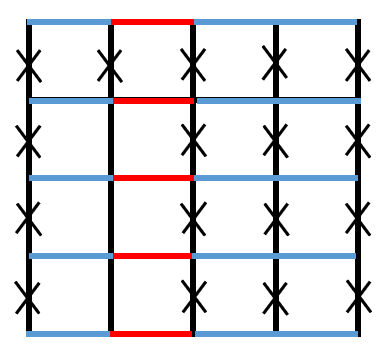}}
  \subfigure[] {\includegraphics[width=0.20 \columnwidth]{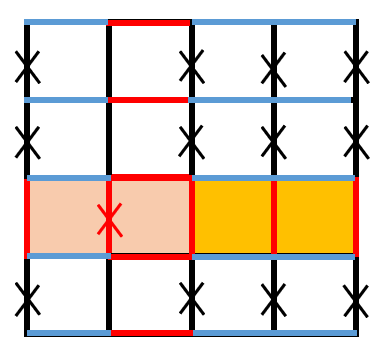}}
  \subfigure[]{\includegraphics[width=0.20 \columnwidth]{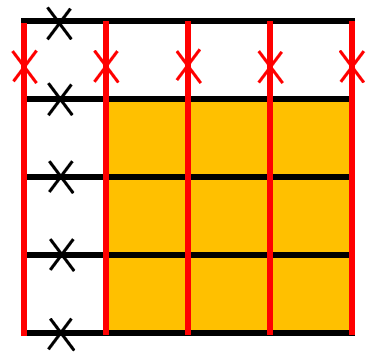}}
  \caption{ Illustration of subsystems in the toric code, and their spanning trees and joint cycles. These subsystems are the qubits on the red edges. (a) The spanning tree of subsystem $A$ is obtained by removing one of the $d$ qubits (red cross) along the non-contractible qubit chain. The spanning tree of subsystem $B$ is obtained by removing appropriate qubits (black cross) in subsystem $B$. The joint cycles are indicated by the orange faces. (b) The spanning forest of subsystem $A$ consists of the qubits on the red edges. 
  The spanning tree of subsystem $B$ is obtained by removing appropriate qubits (black cross) in subsystem $B$. The qubits of $B$ on the blue edges form non-contractible joint cycles with the edges in the spanning forest of $A$. (c) The red edges, with the exception of the one that has been removed (red cross), constitute the spanning forest of subsystem $A$. By removing the edges indicated by the black cross, one can obtain the spanning tree of subsystem $B$. The spanning forest of $A$ and the spanning tree of $B$ share the stabilizer operators that are indicated by one pink face and $d-2$ orange faces. The red edges and the blue edges also form non-contractible joint cycles. (d) The spanning forest of subsystem $A$ and that of subsystem $B$ form a joint cycle space. Two spanning forests share the stabilizer operators which are indicated by the orange faces.}
  \label{fig:example to calculate}
\end{figure}

We calculate the entanglement entropy for several types of subsystem for the logical state $|00\rangle_L$, including a single qubit, two qubits, the qubit chain, the vertical qubit ladder, the cross and all vertical qubits~\cite{Hamma2005bipartie}. Some of these subsystems are shown in Fig.~\ref{fig:example to calculate}.

\subsection{One qubit}

Suppose that subsystem $A$ contains a single qubit, which participates in two $Z$-type stabilizer generators. The boundary of subsystem $A$ contains only one cycle. After deleting a qubit from the boundary to remove the cycle, the qubit in subsystem $A$ and the spanning tree of subsystem $B$ form one joint cycle. Therefore, the entanglement entropy is 1. 

\subsection{Two qubits}

Suppose that subsystem $A$ contains two qubits. These two qubits belong to either a single $Z$-type stabilizer generator, or two distinct $Z$-type stabilizer generators. In the former case, subsystem $A$ and its boundary form 3 stabilizer generators, and the boundary contains one cycle. Therefore, the entanglement entropy is 2. In the latter case, subsystem $A$ and its boundary form 4 stabilizer generators, and the boundary contains two cycles. Consequently, the entanglement entropy is 2.

\subsection{The qubit chain}

Consider a $d \times d$ square lattice and a non-contractible qubit chain in the torus, as shown by the red edges in Fig.~\ref{fig:example to calculate} (a). The spanning tree of subsystem $A$ and that of subsystem $B$ form a joint cycle space with dimension $d-1$. Therefore, the entanglement entropy for the non-contractible qubit chain is $d-1$.

\subsection{The vertical qubit ladder}

Suppose that the qubits in subsystem $A$ form a vertical ladder, as shown by the red edges in Fig.~\ref{fig:example to calculate} (b). The qubits in the vertical ladder are disconnected, forming multiple spanning trees, which is known as the spanning forest. The spanning forest of subsystem $A$ and the spanning tree of subsystem $B$ form $d$ parallel non-contractible joint cycles. Therefore, the entanglement entropy is $d$.

\subsection{The cross}

The qubits in subsystem $A$ form a cross, as shown by the red edges in Fig.~\ref{fig:example to calculate} (c). The spanning forest of subsystem $A$ and the spanning tree of subsystem $B$ form a joint cycle space. A basis of the joint cycle space consists of $d-1$ contractible cycles, represented by a pink face and orange faces, along with $d$ non-contractible cycles.The cyclomatic number of the cycle space is given by $2d-1$, which directly corresponds to the entanglement entropy of the system. Hence, the entanglement entropy is $2d-1$.

\subsection{All vertical qubits}

The subsystem $A$ contains all vertical qubits in the torus, as shown by the red edges in Fig.~\ref{fig:example to calculate} (d). By removing all the vertical edges in the first row, one can obtain the spanning forest of $A$. Similarly, the spanning forest of $B$ can be obtained by removing all horizontal edges from the first column. The spanning forest of $A$ and that of $B$ form a cycle space with dimension $(d-1)^2$, as shown in Fig.~\ref{fig:example to calculate} (d). The entanglement entropy is $(d-1)^2$.

\section{Transformation of parity check matrix}\label{app:transformation-H}

\begin{figure*}[htbp]
\centering
\includegraphics[width=1.0 \columnwidth]{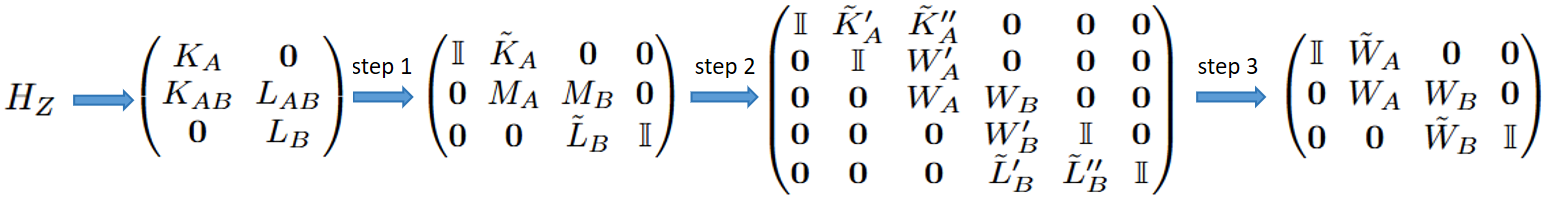}
\caption{The process for transforming the parity check matrix. Prior to step 1, appropriate row and column permutations are applied to reorganize the parity check matrix. Step 1 involves deleting the cycles within subsystem $A$ that do not intersect with the boundary of subsystem $B$, as well as the cycles within subsystem $B$ that do not intersect with the boundary of subsystem $A$. Step 2 focuses on eliminating the cycles that lie on the boundaries of both subsystems $A$ and $B$. Step 3 entails transforming the parity check matrix into its final form through Gaussian elimination.  }
\label{fig:process}
\end{figure*}

We provide a detailed process of transforming the parity check matrix $H_Z$, as summarized in Fig.~\ref{fig:process}.
Once subsystems $A$ and $B$ are defined, the parity check matrix $H_Z$ of a CSS code can be partitioned into $(H_A ~ H_B)$ through appropriate column permutations, where $H_A$ and $H_B$ are submatrices whose columns correspond to qubits in subsystems $A$ and $B$, respectively. By further appropriate row permutations, the parity check matrix can be transformed into 
\begin{eqnarray}
    \begin{pmatrix}
        K_A & \boldsymbol{0}  \\
        K_{AB} & L_{AB} \\
        \boldsymbol{0} & L_B
    \end{pmatrix},
\end{eqnarray}
where the submatrix $(K_A ~ \boldsymbol{0})$ represents stabilizer generators acting non-trivially on qubits only in subsystem $A$, $(\boldsymbol{0} ~ L_B)$ represents stabilizer generators acting non-trivially on qubits only in subsystem $B$, and $(K_{AB} ~ L_{AB})$ represents stabilizer generators acting non-trivially on qubits both in subsystems $A$ and $B$.

For a specific stabilizer generator acting solely on qubits within subsystem $A$, one selects an arbitrary qubit associated with the stabilizer for deletion.
Before deleting the qubit, one can employ Gaussian elimination to eliminate all non-zero elements in the column corresponding to this qubit, with the exception of the entry in the row representing the stabilizer generator.
By further applying appropriate row and column permutations, the parity check matrix can be transformed into
\begin{eqnarray}\label{eq:parity check 1}
    \begin{pmatrix}
        1&K_A'&\boldsymbol{0}\\
        \boldsymbol{0}&\bar{K}_A&\boldsymbol{0}\\
        \boldsymbol{0}&K_{AB}'&L_{AB}\\
        \boldsymbol{0}&\boldsymbol{0}&L_B
    \end{pmatrix},
\end{eqnarray}
where the first column indicates which qubit should be removed. For another stabilizer generator acting only on qubits in subsystem $A$, one applies the same procedure. The parity check matrix in Eq.~(\ref{eq:parity check 1}) can be transformed into
\begin{eqnarray}
    \begin{pmatrix}
        1&0&K_A''&\boldsymbol{0}\\
        0&1&\bar{K}_A'&\boldsymbol{0}\\
        \boldsymbol{0}&\boldsymbol{0}&\bar{K}_A''&\boldsymbol{0}\\
        \boldsymbol{0}&\boldsymbol{0}&K_{AB}''&L_{AB}\\
        \boldsymbol{0}&\boldsymbol{0}&\boldsymbol{0}&L_B
    \end{pmatrix}.
\end{eqnarray}
By repeating the above procedure to  subsystems $A$ and $B$, the parity check matrix can be transformed into the following form,
\begin{eqnarray}\label{eq:matrix form 1}
    \begin{pmatrix}
        \mathbb{I} & \tilde{K}_A & \boldsymbol{0} & \boldsymbol{0} \\
        \boldsymbol{0} & M_A & M_B & \boldsymbol{0} \\
        \boldsymbol{0} & \boldsymbol{0} & \tilde{L}_B &\mathbb{I}
    \end{pmatrix}.
\end{eqnarray}
The columns associated with the two identity matrices correspond to qubits that are needed to be deleted to remove cycles in subsystems $A$ and $B$. The submatrix $M_{AB} \equiv (M_A ~ M_B)$ represents stabilizer generators shared by subsystems $A$ and $B$.

The boundary of subsystem $A$, namely, the qubits associated with the partition of stabilizer generators $M_B$, may contain one or more cycles, which can be generated by multiplying several stabilizer generators shared by $A$ and $B$. This implies that one can always define new stabilizer generators from $M_{AB}$, which act on qubits only in the boundary of subsystem $A$. If such a new generator has been found, we then apply the Gaussian elimination procedure and appropriate row and column permutations to $M_{AB}$, so that it can be transformed into
\begin{eqnarray}
    \begin{pmatrix}
        M_A'&M_B'& \boldsymbol{0} \\
        \boldsymbol{0} &\tilde{M}_B&1
    \end{pmatrix}.
\end{eqnarray}
The new stabilizer generator, represented by $\begin{pmatrix}
    \boldsymbol{0} & \tilde{M}_B&1
\end{pmatrix}$, and the qubit intended to be deleted have been moved to the last row and column, respectively. Continue this process to the submatrix $\begin{pmatrix}
    M_A'&M_B'
\end{pmatrix}$, one can find all the cycles in the boundary $\partial A$. The parity check matrix can be transformed into
\begin{eqnarray}
    \begin{pmatrix}
        M_A'' & M_B'' & \boldsymbol{0} \\
        \boldsymbol{0} & W_B' & \mathbb{I}
    \end{pmatrix}. 
\end{eqnarray}
The submatrix
$\begin{pmatrix}
    \boldsymbol{0} &W_B' & \mathbb{I}
\end{pmatrix}$
represents the stabilizer generators in the boundary of subsystem $A$.

The boundary of subsystem $B$, namely, the qubits associated with the partition of stabilizer generators $M_A$, may also contain cycles, which can be generated by multiplying some stabilizer generators represented by the submatrix $\begin{pmatrix}
    M_A''&M_B''
\end{pmatrix}$. This implies that one can define new stabilizer generators from $\begin{pmatrix}
    M_A''&M_B''
\end{pmatrix}$, which act on qubits only on the boundary of $B$. Following the similar procedure as above, the matrix $M_{AB}$ can be transformed into $W_{AB}$,
\begin{eqnarray}
    W_{AB}
        = \begin{pmatrix}
            \mathbb{I} & W_A' & \boldsymbol{0} & \boldsymbol{0} \\
            \boldsymbol{0} & W_A & W_B & \boldsymbol{0} \\
            \boldsymbol{0} & \boldsymbol{0} & W_B' & \mathbb{I}
        \end{pmatrix}.
\end{eqnarray}
The submatrix 
 $\begin{pmatrix}
     \mathbb{I} & W_A' & \boldsymbol{0} & \boldsymbol{0}
 \end{pmatrix}$
 represents the stabilizer generators enclosed fully in the boundary of $B$. The columns which have only one nonzero element correspond to the qubits needed to be deleted. All cycles have been removed after deleting appropriate qubits in the boundaries of $A$ and $B$, and the remaining stabilizer generators are represented by the submatrix $\begin{pmatrix}
     W_A& W_B
 \end{pmatrix}$.

At this stage, the original parity check matrix $H_Z$ has been transformed into the following form,
\begin{eqnarray}
    \begin{pmatrix}
    \mathbb{I} & \tilde{K}_A^\prime & \tilde{K}_A^{\prime \prime} & \boldsymbol{0} & \boldsymbol{0} & \boldsymbol{0} \\
    \boldsymbol{0} & \mathbb{I} & W_A' & \boldsymbol{0} & \boldsymbol{0} & \boldsymbol{0} \\
    \boldsymbol{0} & \boldsymbol{0} & W_A & W_B & \boldsymbol{0} & \boldsymbol{0} \\
    \boldsymbol{0} & \boldsymbol{0} & \boldsymbol{0} & W_B' & \mathbb{I} & \boldsymbol{0} \\
    \boldsymbol{0} & \boldsymbol{0} & \boldsymbol{0} & \tilde{L}_B' & \tilde{L}_B'' & \mathbb{I}
    \end{pmatrix}.
 \end{eqnarray}
 where we have partitioned the submatrix $\tilde{K}_A$ into $(\tilde{K}_A^\prime ~ \tilde{K}_A^{\prime \prime})$, and the submatrix $\tilde{L}_B$ into $(\tilde{L}_B^\prime ~ \tilde{L}_B^{\prime \prime})$.
 By using the rows in the submatrix $\begin{pmatrix}
    \boldsymbol{0} &\mathbb{I} & W_A' & \boldsymbol{0} & \boldsymbol{0} & \boldsymbol{0}
\end{pmatrix}$ to eliminate the non-zero elements in the block $\tilde{K}_A'$, and the rows in the submatrix $\begin{pmatrix}
    \boldsymbol{0} & \boldsymbol{0} & \boldsymbol{0} & W_B' & \mathbb{I} & \boldsymbol{0}
\end{pmatrix}$ to eliminate the non-zero elements in the block $\tilde{L}_B''$, the parity check matrix can be transformed into
\begin{eqnarray}\label{eq:Hzmodify}
\tilde{H}_Z = 
    \begin{pmatrix}
         \mathbb{I} & \tilde{W}_A & \boldsymbol{0} & \boldsymbol{0} \\
         \boldsymbol{0} & W_A & W_B & \boldsymbol{0} \\
         \boldsymbol{0} & \boldsymbol{0} & \tilde{W}_B & \mathbb{I}
     \end{pmatrix}.
\end{eqnarray}

\section{Proof of the formula for entanglement entropy}\label{app:proof}

After appropriate row and column permutations, the parity check matrix $H_Z$ can be partitioned into 
$\begin{pmatrix}
    H_A&H_B
\end{pmatrix}$. Suppose that $\text{rank}(H_A)=r_A$, $\text{rank}(H_B)=r_B$ and $\text{rank}(H_Z)=r_H$. As a first step, we transform the parity check matrix $H_Z$ via Gaussian elimination to the form that contains $r_H$ linearly independent rows. 
We then implement step 1 in Fig.~\ref{fig:process} and obtain a matrix as shown in Eq.~\eqref{eq:matrix form 1}. 
Suppose that the size of the identity matrix in the first row block in the matrix~\eqref{eq:matrix form 1} is $m_A$, and that of the identity matrix in the third row block is $m_B$, then the rank of $M_{AB}$ is given by
\begin{eqnarray}\label{eq:rank of AB}
    r_{AB} = r_H-m_A-m_B.
\end{eqnarray}
After detecting all cycles in the boundary of subsystem $A$, the submatrix $M_{AB}$ is transformed into
\begin{eqnarray}\label{eq:matrix form 2}
    M_{AB}''
        = \begin{pmatrix}
            M_A'' & M_B'' & \boldsymbol{0} \\
            \boldsymbol{0} & W_B' & \mathbb{I}
        \end{pmatrix}. 
\end{eqnarray}
The rank of the submatrix $(M_A''~ M_B'')$ is equal to $\text{rank}(M_A)$, which is denoted as $r_A''$. 

Since $M_A''$ is a full rank submatrix, we have
\begin{eqnarray}\label{eq:rank equation 1}
    r_A''+m_A=r_A.
\end{eqnarray}
The rank of $M_B''$ is denoted as $r_B''$. Since the number of rows in the identity block in Eq.~(\ref{eq:matrix form 2}) is $r_{AB}-r_A''$, we have
 \begin{eqnarray}\label{eq:rank relation 2}
     r_B''+r_{AB}-r_A''+m_B=r_B.
 \end{eqnarray}
After step 2 in Fig.~\ref{fig:process}, the submatrix $\begin{pmatrix}
    M_A''& M_B''
\end{pmatrix}$ is transformed into
\begin{eqnarray}
    \begin{pmatrix}
        \mathbb{I}&W_A'&\boldsymbol{0}\\
        \boldsymbol{0}&W_A&W_B
    \end{pmatrix}.
\end{eqnarray}
The entanglement entropy is determined by the rank of the submatrix $(W_A~W_B)$, which is equal to the rank of $M_B''$. According to Eqs.~\eqref{eq:rank of AB},~\eqref{eq:rank equation 1} and \eqref{eq:rank relation 2}, $r_B''$ is given by $r_A+r_B-r_H$. Consequently, the entanglement entropy is
\begin{eqnarray}
    S_A=r_A+r_B-r_H.
\end{eqnarray}

\section{Entanglement entropy for logical states}\label{app:logical-state}

When deriving the reduced density matrix and entanglement entropy for general CSS codes, we consider constraints introduced only by stabilizer generators and assume the encoded state to be an equal superposition of all allowable qubit configurations. This is equivalent to considering a state that is an equal superposition of all logical computational bases, namely, 
\begin{equation}
    \ket{\psi_L}= \frac{1}{2^{k/2}} \sum\limits_{C\in \{0, 1\}^{k}}\ket{C}_L,
\end{equation}
where $\ket{C}_L$ represented a logical computational basis state, and $k$ denotes the number of logical qubits.
In fact, our method can also be used to evaluate the entanglement entropy for an equal superposition of a subset of logical computational bases. Denote the logical Pauli $Z$ operators as $\bar{Z}_1, \bar{Z}_2, \cdots, \bar{Z}_{k}$, and the corresponding values of logical qubits as $\bar{z}_1, \bar{z}_2, \cdots, \bar{z}_k$, with $\bar{z}_i = \{0, 1\}$. If one of the logical qubits takes a specific value, say $\bar{z}_1 = 0$, then the encoded state is subject to an additional constraint. By appending a row representing the logical operator $\bar{Z}_1$ to the parity check matrix $H_Z$, we can evaluate the entanglement entropy for the state
\begin{equation}
    \ket{\psi_{L}^{(1)}}= \frac{1}{2^{(k-1) /2}}
    \sum_{\bar{z}_2} \sum_{\bar{z}_3} \cdots \sum_{\bar{z}_k} \ket{0,\bar{z}_2, \cdots, \bar{z}_{k}}_L.
\end{equation}
Similarly, if a set of logical Pauli $Z$ operators take specific eigenvalues, we can calculate the entanglement entropy by adding rows representing these logical Pauli $Z$ operators to the parity matrix $H_Z$ and then apply the graph-based approach.  

If the tensor product (in the code subspace) of two logical Pauli $Z$ operators, say $\bar{Z}_1 \otimes \bar{Z}_2$, takes a specific eigenvalue, say $+1$, then the encoded state is subject to an additional constraint. By appending a row representing the logical operator $\bar{Z}_1 \otimes \bar{Z}_2$ to the parity matrix $H_Z$, we can evaluate the entanglement entropy for the state
\begin{eqnarray}
    \ket{\psi_{L}^{(12)}} &=& \frac{1}{2^{(k-1) /2}}
    \bigg\{ \sum_{\bar{z}_3} \sum_{\bar{z}_4} \cdots \sum_{\bar{z}_k} \ket{0,0, \bar{z}_3, \cdots, \bar{z}_k}_L \nonumber \\
    && + \sum_{\bar{z}_3} \sum_{\bar{z}_4} \cdots \sum_{\bar{z}_k} \ket{1,1, \bar{z}_3, \cdots, \bar{z}_k}_L \bigg\}.
\end{eqnarray}
Similarly, if a set of logical operators, which are in the form of a tensor product of some logical Pauli $Z$ operators, takes specific eigenvalues, we can calculate the entanglement entropy by adding rows representing these logical operators to the parity check matrix $H_Z$ and then apply the graph-based approach.

In the special case where every logical Pauli $Z$ operator takes a specific eigenvalue, we actually evaluate the entanglement entropy for a logical computational basis state $\ket{C}_L$. It can be proved that the entanglement entropy for any logical basis state is the same as that of the logical basis state $\ket{00 \cdots 0}_L$, for which the eigenvalues of all logical Pauli $Z$ operators are equal to one \cite{kargarian2008entanglement}. This equivalence arises from the fact that any logical basis state $\ket{C}_L$ can be transformed into $\ket{00 \cdots 0}_L$ by applying some logical Pauli $X$ operators to flip the corresponding logical qubits from $0$ to $1$, and any logical Pauli $X$ operator is a tensor product of single-qubit Pauli $X$ operators.

Now the question is how to find the logical Pauli $Z$ operators for general CSS codes. We introduce an algorithm to determine the logical Pauli $Z$ operators for an arbitrary CSS code. The algorithm 
starts with the set of $m$ Pauli generators, denoted as $\{g_1, g_2, \cdots, g_m\}$, and proceeds as follows~\cite{wilde2009logical}:
\begin{enumerate}
     \item If the generator $g_1$ commutes with all other generators, set it aside in a “set of processed operators”.
    \item If the generator $g_1$ anti-commutes with another generator
$g_j$, modify
the remaining generators as follows:
\begin{equation*}
\begin{aligned}
     &  \forall i \in \{2, \cdots, m\}, i\neq j
\\
    g_i\to &\left\{\begin{array}{ccc}
        g_i, &[\;g_i, g_1\,]=0,\ &[\;g_i, g_j\,]=0;   \\
        g_i, &\{g_i, g_1\}=0,\ &\{g_i, g_j\}=0;  \\ 
        g_ig_1,& [\;g_i, g_1\,]=0 ,\ &\{g_i, g_j\}=0;  \\
        g_ig_j,& \{g_i, g_1\}=0,\ &[\;g_i, g_j\,]=0.
    \end{array}\right.
\end{aligned}
\end{equation*}
After the modification, all other generators commute with both $g_1$ and $g_j$. Then set $g_1$ and $g_j$ aside in the “set of processed operators”.
\item Execute the above procedure recursively to the remaining generators.
\end{enumerate}
For an arbitrary CSS code, we should first compute the generator matrix $G_Z$ from the parity check matrix $H_Z$. The procedure can be found in Ref.~\cite{nielsen2010quantum}. By applying the algorithm to the generator matrix $G_Z$, where each row represents an independent generator of the code, the output is a new matrix. By removing the rows of the output that are linearly dependent on the rows in $H_Z$, we obtain rows that represent the logical Pauli $Z$ operators.

\section{Subsystem selection}\label{app:subsystems}

In Fig.~\ref{fig:entropy} (a) we compute the von Neumann entropy for a special series of subsystems with increasing sizes. The selection of these subsystems follows a specific procedure, which we describe in detail below. 
\begin{enumerate}
    \item Begin by randomly selecting a stabilizer generator and adding it to a ``waiting set". Initially, subsystem $A$ contains no qubits.
    \item For each stabilizer in the ``waiting set”, append all qubits that are acted on non-trivially by the stabilizer to subsystem A, then apply the graph-based method to compute the von Neumann entropy of the updated subsystem A. Note that subsystem $A$ progressively expands as the procedure iterates through each stabilizer in the "waiting set," gradually incorporating additional qubits.

    \item If subsystem $A$ is smaller than half of the entire system, expand subsystem $A$ via step 4.

    \item For each qubit in subsystem $A$, add all stabilizers that acts non-trivially on it to a new set $\mathbb{S}$. If a stabilizer is already present in the old ``waiting set", do not add it again. Finally, set $\mathbb{S}$ to the ``waiting set", then repeat steps 2 and 3.
\end{enumerate}
By following this procedure, the subsystem expands by adding the qubits that are acted on non-trivially by the adjacent stabilizer. 

\section{Construction of BB codes}\label{app:construction}

We briefly review the construction of the parity check matrices $H_Z$ and $H_X$ for the BB code~\cite{bravyi2024high}. The BB code is a special type of qLDPC codes, for which each stabilizer generator acts non-trivially on six qubits and each qubit participates in six stabilizer generators. The parity check matrices of the BB code are defined using eight parameters, which are denoted as $l, m, a, b, c, d, e, f$. The check matrices are given by~\cite{bravyi2024high}
\begin{equation}
    H_X=[A|B],\quad \quad \quad H_Z=[B^T|A^T],
\end{equation}
where 
\begin{equation}
    A=x^a+x^b+x^c,\quad B=y^d+y^e+y^f.
\end{equation}
Here and below the addition and multiplication of binary matrices
is performed modulo two. The matrices $x$ and $y$ are defined as
\begin{equation}
    x=S_l\otimes I_m,\quad y=I_l\otimes S_m,
\end{equation}
where $I_m$ and $S_m$ is the identity matrix and the cyclic shift matrix of size $m\times m$, respectively. The $i$-th row of $S_m$ has an entry of $1$ in the column $i+1$ (mod $m$).
In our calculation, we consider several BB codes with size up to 756 qubits~\cite{bravyi2024high}. 
The choice of values for the eight parameters is listed in Table.~\ref{tab:parameter}. 

\begin{table}[htbp]
    \centering
    \caption{List of parameters for the construction of various BB codes in our calculation. }
    \begin{tabular}{c|c|c|c|c|c|c|c|c}
    \hline 
        $[[n, k, d]]$ & \ \ $l$ \ \ & \ \ $m$ \ \ & \ \ $a$\ \ & \ \ $b$ \ \ & \ \ $c$\ \ & \ \ $d$ \ \ & \ \ $e$ \ \ & \ \ $f$ \ \  \\
        \hline \hline
         $[[72, 12, 6]]$ & 6 & 6  & 3 & 1 & 2 & 3 & 1 & 2 \\ \hline
        $[[90, 8, 10]]$ & 15 & 3 & 9 & 1 & 2 & 1 & 2 & 7 \\ \hline
        $[[108, 8, 10]]$ & 9 & 6 & 3 & 1 & 2 & 3 & 1 & 2 \\ \hline
        $[[144, 12, 12]]$ & 12 & 6 & 3 & 1 & 2 & 3 & 1 & 2 \\ \hline
        $[[288, 12, 18]]$ & 12 & 12  & 3 & 2 & 7 & 3 & 1 & 2 \\ \hline
        $[[360, 12, \le 24]]$ & 30 & 6  & 9 & 1 & 2 & 3 & 25 & 26 \\ \hline
        \ $[[756, 16, \le 34]]$ \ & 21 & 18  & 3 & 10 & 17 & 5 & 3 & 19 \\ \hline
    \end{tabular}
    \label{tab:parameter}
\end{table}

\section{Construction of quasi cyclic codes}

We summarize the construction of the parity check matrices $H_Z$ and $H_X$ for the quasi cyclic codes \cite{hagiwara2007quantum}. The parity check matrices of the quasi cyclic codes are characterized by three parameters $P$, $\sigma$ and $\tau$, which should have to satisfy certain constraints.

$P$ is an integer. All integers that are coprime to $P$ in the integer module $Z_P$ constitute an abelian group, denoted as $Z_P^*$. $\sigma$ is one of the elements in $Z_P^*$, and $r$ is the order of $\sigma$ such that $\sigma^i-1$ are coprime to $P$ for $1\leq i < r$. $\tau$ is also an element in $Z_P^*$, but is not equal to any power of $\sigma$. The model matrices $C$ and $D$, which will later be used to construct the parity check matrices $H_Z$ and $H_X$, are defined as
\begin{eqnarray}\label{eq:construction of QC model 1}
    C_{ij} = \sigma^{-i+j},~~ \text{for} ~~ 0 \leq j< \frac{L}{2};  ~~ C_{ij}=\tau \sigma^{-i+j}, ~~ \text{for} ~~\frac{L}{2}\leq j<L;
\end{eqnarray}
and 
\begin{eqnarray}\label{eq:construction of QC model 2}
D_{ij} = -\tau\sigma^{i-j}, ~~ \text{for} ~~ 0\leq j< \frac{L}{2}; ~~ D_{ij}= -\sigma^{i-j} ~~ \text{for} ~~\frac{L}{2}\leq j< L;
\end{eqnarray}
where $L = 2r$ is twice the order of $\sigma$. The number of rows in the model matrices $C$ and $D$ can be chosen as an integer that is not greater than $r$. 

The parity check matrices are now constructed by replacing every element $C_{ij}$ of the model matrix $C$, and $D_{ij}$ of the model matrix $D$ by some power of a circulant permutation matrix of size $P$. Suppose that $S_P$ is the circulant permutation matrix, which has an entry of one in the $i$-th row and the $(i+1)$-th column (mod $P$), and has zero entries in the other. The parity check matrix $H_Z$ is formed by substituting every $C_{ij}$ by a submatrix $S_P^{C_{ij}}$, and  $H_X$ is formed by substituting every $D_{ij}$ by a submatrix $S_P^{D_{ij}}$. As an example, we consider $(P,\sigma,\tau) = (7, 2, 5)$ and the order of $\sigma$ is $3$. The circulant permutation matrix $S_7$ is given by 
\begin{eqnarray}
S_7 = 
    \begin{pmatrix}
        0 & 1 & 0 & 0 & 0 & 0 & 0 \\
        0 & 0 & 1 & 0 & 0 & 0 & 0 \\
        0 & 0 & 0 & 1 & 0 & 0 & 0 \\
        0 & 0 & 0 & 0 & 1 & 0 & 0 \\
        0 & 0 & 0 & 0 & 0 & 1 & 0 \\
        0 & 0 & 0 & 0 & 0 & 0 & 1 \\
        1 & 0 & 0 & 0 & 0 & 0 & 0
    \end{pmatrix}. 
\end{eqnarray}
The model matrix $C$ and the corresponding parity matrix $H_Z$ are given by 
\begin{eqnarray}
C = 
    \begin{pmatrix}
        1 & 2 & 4 & 5 & 3 & 6 \\
        4 & 1 & 2 & 6 & 5 & 3 \\
        2 & 4 & 1 & 3 & 6 & 5
    \end{pmatrix}
~~\Longrightarrow ~~
H_Z = 
    \begin{pmatrix}
        S_7 & S_7^2 & S_7^4 & S_7^5 & S_7^3 & S_7^6 \\
        S_7^4 & S_7 & S_7^2 & S_7^6 & S_7^5 & S_7^3 \\
        S_7^2 & S_7^4 & S_7 & S_7^3 & S_7^6 & S_7^5
    \end{pmatrix}.
\end{eqnarray}
The model matrix $D$ and the corresponding parity matrix $H_X$ are given by 
\begin{eqnarray}
D = 
    \begin{pmatrix}
        2 & 1 & 4 & 6 & 3 & 5 \\
        4 & 2 & 1 & 5 & 6 & 3 \\
        1 & 4 & 2 & 3 & 5 & 6
    \end{pmatrix}
~~\Longrightarrow ~~
H_X = 
    \begin{pmatrix}
        S_7^2 & S_7 & S_7^4 & S_7^6 & S_7^3 & S_7^5 \\
        S_7^4 & S_7^2 & S_7 & S_7^5 & S_7^6 & S_7^3 \\
        S_7 & S_7^4 & S_7^2 & S_7^3 & S_7^5 & S_7^6
    \end{pmatrix}.
\end{eqnarray}
From the above procedure to construct the parity check matrices, it is evident that the number of qubits of a QC code is $n = L P$. There is no a close analytic form for the number of encoded logical qubits $k$ for general QC codes. Therefore, $k$ has to be found by calculating the number of independent stabilizers, which can be achieved by applying Gaussian elimination to the parity matrices $H_Z$ and $H_X$. It turns out that for the QC codes considered in this work, the number of encoded logical qubits $k$ can be expressed as  \cite{hagiwara2007quantum}
\begin{eqnarray}
    k=LP-(JP+KP-J-K+2),
\end{eqnarray}
where $J$ and $K$ are the number of rows in $H_Z$ and $H_X$, respectively. The code distance $d$ is estimated by randomly selecting two codewords that are determined by $H_Z$, calculating their Hamming distance, and choosing the minimal Hamming distance as the upper bound of $d$. For the QC codes considered in this work, we have randomly selecting $10^5$ pairs of codewords generated by $H_X$, which constitutes only a subset of all codewords, to estimate the code distance $d$. Due to the large number of codewords, the estimated upper bound for $d$ is looser for larger codes. The QC codes and their corresponding parameters are listed in Table~\ref{tab:choice of QC}.

\renewcommand{\arraystretch}{1.2} 
\setlength{\tabcolsep}{2.5pt}\textbf{}

\begin{table}[htbp]
    \centering
    \caption{List of parameters for the QC codes considered in this work.}
    \begin{tabular}{c|c|c|c|c|c|c}
    \hline 
     [[$n,k,d$]] & \ \ $P$ \ \ & $\ \ \sigma$ \ \ &\ \  $\tau$ \ \ &\ \ $r$\ \ &\ \ $J$\ \ &\ \ $K$\ \ \\
    \hline    
    \hline
        [[$42,4,\leq6$]] &$13$&$3$&$2$&$3$&$3$&$3$\\
    \hline
        [[$78,4,\leq14$]]&$13$&$3$&$2$&$3$&$3$&$3$ \\
    \hline
        [[$114,4,\leq24$]]&$19$&$7$&$2$&$3$&$3$&$3$ \\
    \hline
        [[$258,4,\leq56$]]&$43$&$6$&$2$&$3$&$3$&$3$ \\
    \hline
        [[$582,4,\leq132$]]&$97$&$35$&$2$&$3$&$3$&$3$\\
    \hline
        [[$104,6\leq14$]]&$13$&$5$&$2$&$4$&$4$&$4$\\
    \hline    
        [[$136,6,\leq24$]]&$17$&$4$&$2$&$4$&$4$&$4$\\
    \hline
        [[$232,6,\leq50$]]&$29$&$12$&$2$&$4$&$4$&$4$\\
    \hline
        [[$424,6,\leq96$]]&$53$&$23$&$2$&$4$&$4$&$4$\\
    \hline  
    \ \ [[$584,6,\leq136$]] \ \ &$73$&$27$&$2$&$4$&$4$&$4$\\
    \hline    
    \end{tabular}
    \label{tab:choice of QC}
\end{table}

\newpage

\section{Entanglement entropy and graph-theoretic quantities}

The entanglement entropy can be expressed in terms of graph-theoretic quantities: the number of vertices shared between subsystems and the numbers of connected components within each subsystem. We consider two cases in which the entire system is connected and disconnected.  

We first consider the case where the entire system is connected. Assume that subsystem $A$ consists of $K_1$ connected components, denoted as $A_i~(i=1, \cdots, K_1)$; and subsystem $B$ consists of $K_2$ connected components, denoted as $B_j~(j=1, \cdots, K_2)$. The set of vertices that is shared between the component $A_i$ and subsystem $B$ is denoted $V_{A_i \cap B}$. Similarly, $V_{B_j\cap A}$ denotes the set of vertices that are shared between the component  $B_j$ and subsystem $A$. These quantities satisfy the relation
\begin{eqnarray}
    \sum_i |V_{A_i\cap B}|=\sum_j |V_{B_j\cap A}|=|V_{A\cap B}|,
\end{eqnarray}
where $V_{A\cap B}$ is the set of vertices shared between subsystems $A$ and $B$, and $|V_{A\cap B}|$ is the number of common vertices.

The spanning tree $T_{A_i}$ is obtained by removing appropriate edges from the connected component $A_i$. Note that the number of vertices of the spanning tree $T_{A_i}$ is the same as that of $A_i$, and the number of edges of $T_{A_i}$ is $|A_i|-1$. The spanning tree/forest $T_A$ of subsystem $A$ is a union of all spanning trees $T_{A_i}$. Since there are no common vertices and edges between different components $A_i$, therefore, the number of vertices and edges of $T_A$ is $|V_A| = \sum_i |V_{A_i}|$ and $\sum_i |V_{A_i}| - K_1$, respectively. Suppose that $T_{B_j}$ is the spanning tree within the component $B_j$, and the spanning tree/forest $T_B$ of subsystem $B$ is the union of all spanning trees $T_{B_j}$. We can similarly find that the number of vertices and edges of $T_B$ is $|V_B| = \sum_j |V_{B_j}|$ and $\sum_j |V_{B_j}| - K_2$, respectively.

The cyclomatic number $\mathbf{c}(G)$ for a connected graph $G$ is given by 
\begin{eqnarray}\label{eq:cyclomatic-number}
    c(G) =|E|-|V|+1,
\end{eqnarray}
where $|E|$ and $|V|$ are the number of edges and vertices in $G$, respectively. The core in developing the graphical formalism for entanglement entropy involves applying Eq.~\eqref{eq:cyclomatic-number} to the joint graph $T_A\cup T_B$. The number of edges of $T_A\cup T_B$ is the sum of the number of edges of $T_A$ and that of $T_B$, namely, 
\begin{eqnarray}
    |E_{T_A\cup T_B}| = \sum _i|V_{A_i}|+\sum_j|V_{B_j}|-K_1-K_2 = |V_A|+|V_B|-K_1-K_2;
\end{eqnarray}
and the number of vertices of $T_A\cup T_B$ is 
\begin{eqnarray}
    |V_{T_A\cup T_B}| = |V_A|+|V_B|-|V_{A\cap B}|,
\end{eqnarray}
where $V_{T_A\cup T_B}$ is the set of common vertices between subsystems $A$ and $B$. The entanglement entropy can be expressed as 
\begin{eqnarray}\label{eq:graphic-form-1}
    S_A=|c(T_A\cup T_B)|=|E_{T_A\cup T_B}| - |V_{T_A\cup T_B}| + 1=|V_{A\cap B}| -K_1-K_2+1.
\end{eqnarray}

In the second case, the entire system is disconnected and consists of $K$ components that are denoted as $\left\{C_i\right\}_{i=1}^K$. In each component $C_i$, suppose that the subgraph within $A$ is partitioned into $K_1^i$ connected components $\left\{A_l^{i}\right\}_{l=1}^{K_1^i}$, and that within $B$ is partitioned into $K_2^i$ connected components $\left\{B_j^i\right\}_{j=1}^{K_2^i}$. The set of vertices that is shared between the component $A_l^i$ and the subsystem $B$ is denoted by $V_{A_l^i\cap B}$ for each $C_i$. Similarly, $V_{B_j^i\cap A}$ denotes the set of vertices that is shared between the component $B_j^i$ and the subsystem $A$ for each $C_i$. These quantities satisfy the following relations:
\begin{eqnarray}
    \sum_l|V_{A_l^i\cap B}| &=& \sum_j|V_{B_j^i\cap A}|=|V_{A\cap B}^i|, \nonumber\\
    \sum_i|V_{A\cap B}^i| &=& |V_{A \cap B}|,
\end{eqnarray}
where $|V_{A\cap B}^i|$ is the number of common vertices shared between subsystems $A$ and $B$ within the component $C_i$, and $|V_{A\cap B}|$ is the number of the total common vertices shared between subsystems $A$ and $B$.

The joint graph $T_A\cup T_B$ can be decomposed as a union of a set of subgraphs,
\begin{eqnarray}
    T_A\cup T_B=\bigcup_i \big( (T_A\cup T_B)\cap C_i \big)=\bigcup_i(T_A\cup T_B)_i,
\end{eqnarray}
where $(T_A\cup T_B)_i= (T_A\cup T_B)\cap C_i$. The cyclomatic number of each subgraph $(T_A\cup T_B)_i$ is given by 
\begin{eqnarray}
    c((T_A \cup T_B)_i)=|V_{A \cap B}^i|-K_1^i-K_2^i+1.
\end{eqnarray}
Summing over all cyclomatic numbers of the subgraphs $(T_A\cup T_B)_i$ gives the total cyclomatic number of the joint graph $T_A\cap T_B$,
\begin{eqnarray}
    c(T_A \cap T_B)=\sum_{i=1}^K (|V_{A\cap B}^i|-K_1^i-K_2^i)+K=|V_{A\cap B}|-\sum_i(K_1^i+K_2^i)+K.
\end{eqnarray}
Recognizing that $\sum_iK_1^i=K_1$ and $K_2^i=K_2$, we obtain the general expression for the entanglement entropy,
\begin{eqnarray}
    S_A=|V_{A\cap B}|-K_1-K_2+K.
\end{eqnarray}
This result suggests that the entanglement entropy depends on both the connection strength between subsystems, quantified by $|V_{A\cap B}|$, and the fragmentation within each subsystem, characterized by their respective numbers of components $K_1$ and $K_2$. In addition, the entanglement entropy is affected by the fragmentation the entire system, as indicated by the term $K$.

\section{Qubit duplication for general CSS codes}

Not every CSS code can be directly represented as a simple graph, as is the case for toric codes. This arises particularly when the column weight of the parity-check matrix exceeds two. We develop a method to modify the parity matrix of a general CSS code such that it can be represented by a simple planar graph. 

\begin{figure*}[htbp]
  \centering
  \includegraphics[width=0.98 \columnwidth]{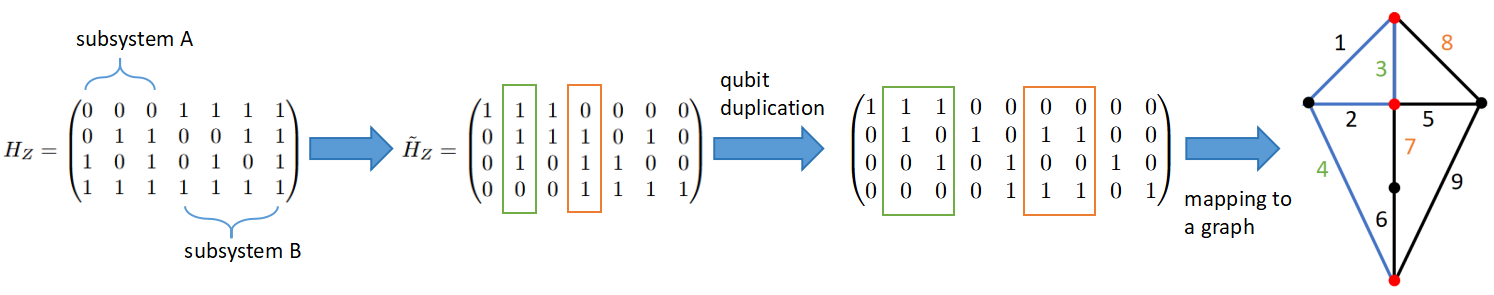}
  \caption{Graph representation of a high-column-weight CSS code via qubit duplication. The parity-check matrix $H_Z$
  is transformed into $\tilde{H}_Z$ as derived in Appendix \ref{app:transformation-H}. Qubit duplication splits columns with weight greater than $2$ into multiple columns of weight at most $2$ (highlighted in green and orange boxes). The duplicated matrix is mapped to a graph where each edge represents a physical qubit, with edge labels denoting the column indices in the duplicated matrix. Duplicated qubits are explicitly indicated: edge $4$ is a duplicate of edge $3$ (green), and edge $8$ is a duplicate of edge $7$ (orange). Qubits in subsystem $A$ are blue, and those in subsystem $B$ are black. The three red vertices are shared between subsystems $A$ and $B$. All subsystems and the entire system are connected $(K_1=K_2=K=1)$. The entanglement entropy $S_A=2$ is obtained from both the rank analysis of $H_Z$ and the graph-theoretic formula $S_A=|V_{A\cap B}|-K_1-K_2+K$, demonstrating consistency between these two methods.}
  \label{fig:replicated process}
\end{figure*}

We first transform the parity-check matrix $H_Z$ into $\tilde{H}_Z$ given by Eq.~\eqref{eq:Hzmodify}, in which the identity blocks correspond to qubits that need to be deleted to remove cycles. For any remaining qubit that participates in more than two stabilizer generators, we split its corresponding column into multiple columns such that each new column has a weight of at most two. Each new column represents a duplicate of the original qubit, implying that it always has the same qubit value as the original qubit. After this duplication procedure, the parity-check matrix becomes
\begin{eqnarray}
    \begin{pmatrix}
        \mathbb{I} & \tilde{W}_A^{(1)}&\tilde{W}_A^{(2)} & \boldsymbol{0} & \boldsymbol{0} &\boldsymbol{0}\\
        \boldsymbol{0} & W_A^{(1)} &W_A^{(2)}& W_B^{(1)}&W_B^{(2)} & \boldsymbol{0} \\
        \boldsymbol{0} & \boldsymbol{0} & \boldsymbol{0}&\tilde{W}_B^{(1)}&\tilde{W}_B^{(2)} &\mathbb{I}
    \end{pmatrix},
\end{eqnarray}
where the columns of $W_A^{(1)}$ and $W_B^{(2)}$ correspond to the duplicated qubits, each now satisfying the column-weight constraint. Crucially, this duplication does not alter the entanglement entropy since $(\boldsymbol{0}~~ W_A^{(1)}~~W_A^{(2)}~~W_B^{(1)}~~W_B^{(2)}~~\boldsymbol{0})$ remains full-rank throughout the process. With all columns now of weight at most two, the parity-check matrix can be directly interpreted as the incidence matrix of a simple graph. We consider the Hamming code $[7,1,3]$ as an example to illustrate this method, as illustrated in Fig.~\ref{fig:replicated process}.

By controlling the number of duplications per qubit, one can construct different graph representations for the same CSS code. The entanglement entropy of a subsystem $A$ is invariant which can then be evaluated using the graph-theoretic formula: $S_A=|V_{A\cap B}|-K_1-K_2+K$. This method provides a systematic way to represent general CSS codes as graphs, enabling the application of graph-based tools to analyze their entanglement structure.

\section{Analysis on rate of change of entropy discrepancy for toric codes}

In this appendix, we analyze in detail the rate of change of the entropy discrepancy for toric codes. By using the expression for the entanglement entropy $S_A$ in terms of the numbers of common vertices between subsystems $A$ and $B$, and their corresponding components, we find
\begin{eqnarray}
   I_A=n_A-\bar{S}_A(n_A)=|E_A|-\overline{|V_{A\cap B}|}+\bar{K}_1+\bar{K}_2-1.
\end{eqnarray}
Here, $|E_A|$ is the number of edges (qubits) in subsystem $A$, $|V_{A\cap B}|$ is the number of vertices in the intersection of subsystems $A$ and $B$, and $K_1$ and $K_2$ are the numbers of connected components in subsystems $A$ and $B$, respectively; the overbar represents an average over randomly selected subsystems. Differentiating the above expression for $I_A$ with respect to $n_A$ yields the rate of change of the entropy discrepancy, 
\begin{eqnarray}
    \frac{d I_A}{d n_A}=1-\frac{d \overline{|V_{A\cap B}|}}{d|E_A|}+\frac{d\bar{K}_1}{d|E_A|}+\frac{d\bar{K}_2}{d|E_A|}.
\end{eqnarray}

In order to qualitatively understand the behavior of the rate of change of the entropy discrepancy, we consider how it changes when transferring one qubit from subsystem $B$ to subsystem $A$. When subsystem $A$ is much smaller than half of the entire code, with high probability, subsystem $A$ consists of multiple connected components and subsystem $B$ is a single connected component. Suppose that we randomly select a qubit from subsystem $B$ and add it to subsystem $A$. There are four main classes of possibilities, which consist of a total of 14 cases, as listed in Table~\ref{tab:n_A is small}. 

In the first class, the selected qubit is not connected to any existing component of $A$. Transferring this qubit to $A$ introduces an additional component to $A$ and two more common vertices. Therefore, the entropy discrepancy remains the same. In the second class, the selected qubit is connected to only one of the existing components of $A$, but does not form any cycle. The entropy discrepancy is found to increase by one when an $X$-type stabilizer is formed by the selected qubit and some of those in the existing component, and remains the same when no $X$-type stabilizer is formed. In the third class, the selected qubit connects two of the existing components of $A$, therefore reducing the number of components of $A$ by one. Similarly, the entropy discrepancy increases by one when an $X$-type stabilizer is formed by the selected qubit and some of those in the existing component, and remains the same when no $X$-type stabilizer is formed. In the fourth class, transferring the selected qubit to $A$ forms a cycle in it. The entropy discrepancy is always found to increase (by one or two) in this class. It is evident that transferring a qubit from $B$ to $A$ increases the entropy discrepancy only when a cycle or an $X$-type stabilizer is formed in subsystem $A$.

\renewcommand{\arraystretch}{1.0} 
\setlength{\tabcolsep}{0.5pt}

\begin{table}[htbp]
    \centering
    \caption{ Change of entropy discrepancy when transferring a qubit from a connected subsystem $B$ to subsystem $A$. Fourteen different cases are classified into four distinct classes based on the impact on subsystem $A$. In class (I), a new component is introduced to subsystem $A$; in class (II), the selected qubit is connected to only one of the existing components of $A$ but does not form any cycle; in class (III), two existing components of subsystem $A$ is merged into one; in class (IV), a new cycle is formed within subsystem $A$. The table provides a list for the degrees $d_A(v_i)$ of the vertices associated with the transferred edge within subsystem $A$, the change in the number of common vertices $\Delta|V_{A\cap B}|$, changes in component counts $\Delta K_1$ and $\Delta K_2$, the change in the entanglement entropy $\Delta S_A$, and the resultant change in the entropy discrepancy $\Delta I_A$. Here $\left\{A_l\right\}$ represent the set of the components in subsystem $A$. }  

\begin{tabular}{|c|c|c|c|c|c|c|c|c|c|}
\hline
~Class~ & ~Case~ & ~Transferred qubit, denoted by $(v_i, v_j)$~ & ~$d_A(v_i)$~ & ~$d_A(v_j)$~ & ~$\Delta |V_{A\cap B}|$~ & $~\Delta K_1$~ & ~$\Delta K_2$~ & ~$\Delta S_A$~  & ~$\Delta I_A$~ \\
\hline

\multirow{1}{*}{I} & $1$&$v_i,v_j\in V_B\setminus V_{A\cap B}$ &$0$& $0$& $2$ & $1$ & $0$ & $1$ &$0$ \\
\hline

\multirow{3}{*}{II} & $2$ &$v_i\in V_{A_l\cap B},v_j\in V_B\setminus V_{A\cap B} $ &$1$&$0$ & $1$ & $0$ & $0$ & $1$ & $0$\\ \cline{2-10}
 & $3$ &$v_i\in V_{A_l\cap B},v_j\in V_B\setminus V_{A\cap B} $  &$2$&$0$& $1$ & $0$ & $0$ & $1$ & $0$ \\ \cline{2-10}
 & $4$ &$v_i\in V_{A_l\cap B},v_j\in V_B\setminus V_{A\cap B}$  &$3$&$0$ & $0$ & $0$ & $0$ & $0$ & $1$ \\   
\hline

\multirow{5}{*}{III} & $5$ & $v_i\in V_{A_l\cap B},v_j\in V_{A_{l'}\cap B}$ &$1$&  $1$ & $0$ & $-1$ & $0$ & $1$ &$0$ \\ \cline{2-10}
 & $6$ & $v_i\in V_{A_l\cap B},v_j\in V_{A_{l'}\cap B}$ &$1$&$2$ & $0$ & $-1$ & $0$ & $1$ & $0$\\ \cline{2-10}
 & $7$ & $v_i\in V_{A_l\cap B},v_j\in V_{A_{l'}\cap B}$ &$2$& $2$ & $0$ & $-1$ & $0$ & $1$ & $0$\\ \cline{2-10}
 & $8$ & $v_i\in V_{A_l\cap B},v_j\in V_{A_{l'}\cap B}$  &$1$&$3$ & $-1$ & $-1$ & $0$ & $0$ & $1$\\ \cline{2-10}
 & $9$ & $v_i\in V_{A_l\cap B},v_j\in V_{A_{l'}\cap B}$ &$2$&$3$  & $-1$ & $-1$ & $0$ & $0$ & $1$\\
\hline

\multirow{5}{*}{IV} & $10$ & $v_i,v_j\in V_{A_l\cap B}$  &$1$&$1$& $0$ & $0$ & $0$ & $0$ & $1$\\ \cline{2-10}
 & $11$ & $v_i,v_j\in V_{A_l\cap B}$ &$1$& $2$& $0$ & $0$ & $0$ & $0$ & $1$\\ \cline{2-10}
 & $12$ & $v_i,v_j\in V_{A_l\cap B}$  &$2$&$2$& $0$ & $0$ & $0$ & $0$ & $1$\\ \cline{2-10}
 & $13$ & $v_i,v_j\in V_{A_l\cap B}$  &$1$&$3$& $-1$ & $0$ & $0$ & $-1$ & $2$\\ \cline{2-10}
 & $14$ & $v_i,v_j\in V_{A_l\cap B}$  &$2$&$3$ & $-1$ & $0$ & $0$ & $-1$ & $2$\\ 
\hline  

\end{tabular}
\label{tab:n_A is small}
\end{table}

When $n_A$ is sufficiently small, the number of edges that are incident to $A$ is significantly smaller than those that are not incident to it. In this regime, class I is dominant when a qubit is randomly selected from $B$ and added to $A$. As the size of $A$ increases, classes II and III become more and more probable, whilst class IV remains less probable. This explains the almost vanishing rate of change of the entropy discrepancy when subsystem $A$ is small. As subsystem $A$ continues to expand, the probability of forming a cycle or an $X$-type stabilizer in $A$ increases, leading to a significant growth of $d I_A/dn_A$.

When subsystem $A$ is much larger than half of the entire code, with high probability, subsystem $A$ is a single connected component, whereas subsystem $B$ typically comprises of multiple connected components. Suppose that we randomly select a qubit from $B$ and add it to $A$. There are four main classes of possibilities, which consist of 14 cases, as listed in Table~\ref{tab:n_A is large}.

In the first class, the selected qubit itself is an existing component of $B$. Transferring this qubit to $A$ reduces one component of $B$ and two common vertices. Therefore, the entropy discrepancy increases by two. In the second class, the selected qubit is in an edge with an endpoint of degree one. The entropy discrepancy is found to increase by one when an $X$-type stabilizer is removed from $B$, and increases by two when no $X$-type stabilizer is removed. In the third class, the selected qubit is a bridge without endpoints, the removal of which breaks one component of $B$ into two components. Similarly, the entropy discrepancy increases by one when an $X$-type stabilizer is removed from $B$, and increases by two when no $X$-type stabilizer is removed. In the fourth class, the selected qubit is not a bridge but with a pair of endpoints, the removal of which always breaks a cycle in $B$. The entropy discrepancy is always found to increase at most by one in this class.

When $n_A$ is sufficiently large, the number of edges that are incident to $B$ is significantly smaller than those that are not incident to it. In this regime, class I is dominant when a qubit is randomly selected from $B$ and added to $A$. This explains the almost constant (approximately equal to 2) rate of change of the entropy discrepancy when subsystem $A$ is close to the entire system.

\renewcommand{\arraystretch}{1.0} 
\setlength{\tabcolsep}{0.5pt}

\begin{table}[htbp]
    \centering
    \caption{Change of entropy discrepancy when transferring a qubit from a multi-component subsystem $B$ to a connected subsystem $A$. Fourteen cases are classified into four distinct classes based on the impact on subsystem $B$. (I) Removal of an isolated component from subsystem $B$, (II) removal of a bridge edge with an endpoint, (III) removal of a bridge edge without any endpoints (disconnecting a component of subsystem $B$), and (IV) breaking a cycle in subsystem $B$. The table provides a list for the degrees $d_B(v_i)$ of the vertices within subsystem $B$, elements in the bridge set $\mathbf{Br}(B_k)$ of the component $B_k$, the change in the number of common vertices $\Delta|V_{A\cap B}|$, changes in component counts $\Delta K_1$ and $\Delta K_2$, the change in entanglement entropy $\Delta S_A$, and the resultant change in the entropy discrepancy $\Delta I_A$. A bridge is an edge that, if deleted, disconnects the graph, separating some vertices from the rest.} 
\begin{tabular}{|c|c|c|c|c|c|c|c|c|c|}
\hline
~Class~ & ~Case~ & ~Transferred qubit, denoted by $(v_i, v_j)$~ & ~$d_B(v_i)$~ & ~$d_B(v_j)$~ & ~$\Delta |V_{A\cap B}|$~ & $ ~\Delta K_1$ & ~$\Delta K_2$~ & $ ~\Delta S_A$~  & ~$\Delta I_A$~ \\
\hline

\multirow{1}{*}{I} & $1$&$(v_i,v_j)\in \mathbf{Br}(B_{k})$  &$1$&$1$& $-2$ & $0$ & $-1$ & $-1$ & $2$\\
\hline

\multirow{3}{*}{II} & $2$&$(v_i,v_j)\in \mathbf{Br}(B_{k})$  &$2$&$1$& $-1$ & $0$ & $0$ & $-1$ & $2$\\ \cline{2-10}
 & $3$&$(v_i,v_j)\in \mathbf{Br}(B_{k})$  &$3$&$1$& $-1$ & $0$ & $0$ & $-1$ & $2$\\ \cline{2-10}
 & $4$&$(v_i,v_j)\in \mathbf{Br}(B_{k})$ &$4$& $1$& $0$ & $0$ & $0$ & $0$ & $1$\\
\hline

\multirow{5}{*}{III} & $5$&$(v_i,v_j)\in \mathbf{Br}(B_{k})$ &$2$& $2$& $0$ & $0$ & $1$ & $-1$ & $2$\\ \cline{2-10}
 & $6$&$(v_i,v_j)\in \mathbf{Br}(B_{k})$  &$3$&$2$ & $0$ & $0$ & $1$ & $-1$ & $2$\\ \cline{2-10}
 & $7$&$(v_i,v_j)\in \mathbf{Br}(B_k)$  &$2$&$4$& $1$ & $0$ & $1$ & $0$ & $1$\\ \cline{2-10}
 & $8$&$(v_i,v_j)\in \mathbf{Br}(B_{k}) $ &$3$&$4$ & $1$ & $0$ & $1$ & $0$ & $1$ \\ \cline{2-10}
 & $9$&$(v_i,v_j)\in \mathbf{Br}(B_{k})$ &$4$&$4$ & $2$ & $1$ & $1$ & $0$ &$1$ \\ 
\hline

\multirow{5}{*}{IV} & $10$&$(v_i,v_j)\notin \mathbf{Br}(B_{k})$ &$2$&$2$ & $0$ & $0$ & $0$ & $0$ &$1$ \\ \cline{2-10}
 & $11$&$(v_i,v_j)\notin \mathbf{Br}(B_k)$ &$3$&$2$ & $0$ & $0$ & $0$ & $0$ & $1$\\ \cline{2-10}
 & $12$&$(v_i,v_j)\notin \mathbf{Br}(B_{k})$ &$2$&$4$ & $1$ & $0$ & $0$ & $1$ &$0$ \\ \cline{2-10}
 & $13$&$(v_i,v_j)\notin \mathbf{Br}(B_{k})$  &$3$& $4$& $1$ & $0$ & $0$ & $1$ & $0$  \\ \cline{2-10}
 & $14$&$(v_i,v_j)\notin \mathbf{Br}(B_{k})$  &$4$& $4$ & $2$ & $1$ & $0$ & $1$ & $0$\\  
\hline 

\end{tabular}
\label{tab:n_A is large}
\end{table}

\end{document}